%% file: asplos24.tex
\documentclass[10pt, conference]{ieeetran}

\usepackage[normalem]{ulem}
\usepackage{braket}
\usepackage{cite}
\usepackage{amsmath,amssymb,amsfonts}
\usepackage{algorithmic}
\usepackage{graphicx}
\usepackage{textcomp}
\usepackage{xcolor}
\usepackage{braket}
\usepackage{algorithmic}
\usepackage{graphicx}
\usepackage{tabularx}

\newcommand*{\sysname}{\textcolor{black}{FDeQ}}
\usepackage{graphicx}
\IEEEoverridecommandlockouts

\usepackage{fancyhdr}
\fancypagestyle{firstpage}{%
  \rhead{Accepted article to appear in the proceedings of IEEE QCE’24}
}

\begin{document}

\title{\huge A Quantum Approximate Optimization Algorithm-based Decoder Architecture for NextG Wireless Channel Codes}
\author{Srikar Kasi$^{\star, \dagger}$, James Sud$^{\star, \ddagger}$\thanks{$\star$ Co-primary authors}, Kyle Jamieson$^\dagger$, Gokul Subramanian Ravi$^{\S}$\\
\IEEEauthorblockA{$^\dagger$Princeton University, $^\ddagger$Univeristy of Chicago, $^\S$University of Michigan}}

\date{}
\maketitle
\thispagestyle{firstpage}

\input{0_abstract}
\input{1_intro}

\input{2_primer_coding}

\input{3_primer_quantum}

\input{4_design}

\input{5_implementation}

\input{6_evaluation}

\input{7_resource_estimation}

\input{8_backend}

\input{acks}

\bibliographystyle{unsrt}
\bibliography{reference}

\end{document}

%% file: 0_abstract.tex
\begin{abstract}

Forward Error Correction (FEC) provides reliable data flow in wireless networks despite the presence of noise and interference. However, its processing demands significant fraction of a wireless network's resources, due to its computationally-expensive decoding process. This forces network designers to compromise between performance and implementation complexity. In this paper, we investigate a novel processing architecture for FEC decoding, one based on the quantum approximate optimization algorithm (QAOA), to evaluate the potential of this emerging quantum compute approach in resolving the decoding performance--complexity tradeoff.

We present \sysname, a \underline{Q}AOA-based \underline{F}EC \underline{De}coder design targeting the popular NextG wireless Low Density Parity Check (LDPC) and Polar codes. To accelerate QAOA-based decoding towards practical utility, \sysname{} exploits temporal similarity among the FEC decoding tasks. This similarity is enabled by the fixed structure of a particular FEC code, which is independent of any time-varying wireless channel noise, ambient interference, and even the payload data. We evaluate \sysname{} at a variety of system parameter settings in both ideal (noiseless) and noisy QAOA simulations, and show that \sysname{} achieves successful decoding with error performance at par with state-of-the-art classical decoders at low FEC code block lengths. Furthermore, we present a holistic resource estimation analysis, projecting quantitative targets for future quantum devices in terms of the required qubit count and gate duration, for the application of \sysname{} in practical wireless networks, highlighting scenarios where \sysname{} may outperform state-of-the-art classical FEC decoders.

\end{abstract}

%% file: 1_intro.tex
\section{Introduction}\label{sec:intro}

As wireless networks continue to evolve, scaling up \textit{spectral efficiency} (measured in bits per second per Hz of the frequency spectrum) has been of paramount importance. In fifth generation (5G) cellular wireless networks, this is achieved through the use of millimeter-wave communication, Multiple Input Multiple Output (MIMO) communication, and densely packed small cell deployments \cite{rappaport2013millimeter, bogale2016massive}. Despite progress in these areas, network designers are already looking ahead to the NextG roadmap, envisioning a future with even greater data usage where NextG networks are expected to handle data rates up to 1,000$\times$ greater than 5G. To scale up spectral efficiency further, NextG systems will consider implementing advanced techniques such as ultra-massive MIMO arrays and robust forward error correction coding schemes \cite{kim2021heuristic}. This will result in exponentially greater computational requirements than current 5G systems. Consequently, there is a pressing need to search for computationally efficient hardware solutions for wireless networks to meet these demands.

\begin{figure}
    \centering \includegraphics[width=\columnwidth]{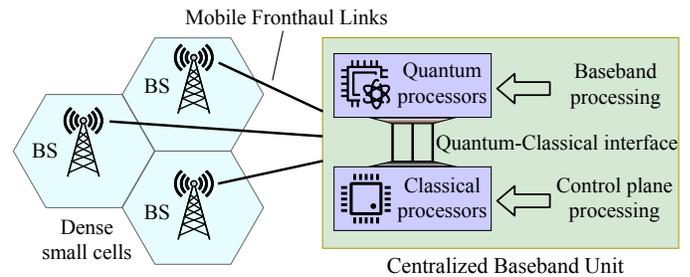}
    \caption{Our envisioned deployment scenario of quantum gate processing based wireless communication in a Centralized RAN context.}
    \label{fig:scenario}
\end{figure}

\begin{figure}
    \centering  \includegraphics[width=0.8\columnwidth]{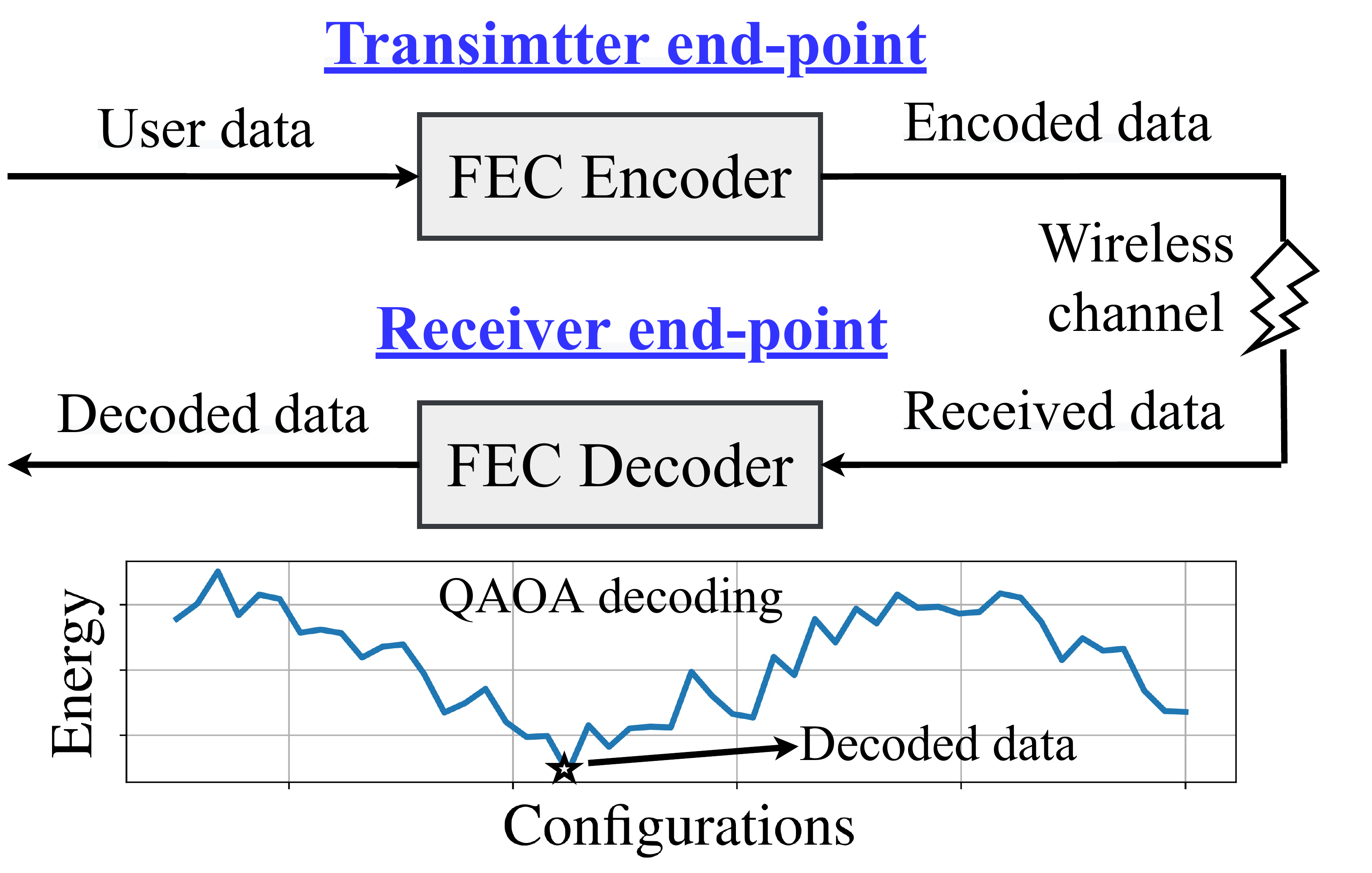}
    \caption{FEC processing architecture in wireless networks. A QAOA-based FEC decoder returns the lowest energy configuration of the optimization problem as the decoded solution.}
    \label{fig:intro_1}
\end{figure}

Figure~\ref{fig:scenario} depicts our envisioned scenario in a wireless Centralized Radio Access Network (C-RAN) context, where quantum gate processors are used for heavyweight baseband tasks and classical processors undertake lightweight control plane tasks \cite{kasi2021cost}. A key component of the baseband processing is the Forward Error Correction (FEC) code \cite{peterson1972error}, a construct that adds redundancy into the data transmission to correct bit errors introduced by noise and interference of the wireless channel. Fig.~\ref{fig:intro_1} shows a typical FEC processing architecture in wireless networks, where a transmitter (\textit{e.g.,} mobile) performs encoding and a receiver (\textit{e.g.,} BS) performs decoding of the FEC code respectively. FEC codes that achieve Shannon capacity exist today,\footnote{Shannon capacity is the upper bound on data transmission rate beyond which reliable communication is not guaranteed \cite{shannon1948mathematical}.} but realizing their potential in practice has been challenging due to their computationally-expensive decoder processing requirements. In particular, traditional FEC decoders entail two major problems: First, to achieve high spectral efficiency, classical FEC decoders must operate at high clock speeds--which has already reached a plateau for CMOS devices \cite{itrs}. Second, the optimal FEC decoding performance comes at the price of exponential computational complexity, which will not scale for a CMOS hardware implementation \cite{xu2007complexity}. These issues thus hinder the utility of traditional FEC decoders in high spectral efficiency networks envisioned in NextG systems. To address these issues, researchers are beginning to investigate alternate approaches to classical FEC decoders \cite{kasi2020towards, kasi2022design, bian2014discrete, chancellor2016direct}. In this spirit, this paper explores the potential of quantum gate computing for the task of FEC decoding, as an alternative to CMOS.

Quantum computing (QC) is a growing computational paradigm, holding great potential for substantial advancements in cryptography~\cite{Shor_1997}, chemistry~\cite{kandala2017hardware}, optimization~\cite{moll2018quantum}, and beyond \cite{biamonte2017quantum}. In the current stage of Noisy Intermediate-Scale Quantum computing~\cite{preskill2018quantum}, and with the advent of the Early Fault Tolerance era on the horizon, the emphasis is on navigating the realm of QC devices characterized by thousands of imperfect qubits, grappling with non-trivial error rates, and gradually progressing towards millions of qubits. As these quantum devices continue to grow in size and improve in qubit quality, a critical objective is to identify practical real-world applications that can leverage the potential of these devices in both the short and long terms. Furthermore, optimizing the entire quantum software-hardware stack is paramount to ensure the practical efficacy of QC devices and to meet the specific demands of target applications.

This work tackles the FEC decoding problem via an iterative hybrid quantum--classical algorithm known as Quantum Approximate Optimization Algorithm or QAOA \cite{farhi2014quantum}. 
QAOA is promising for FEC decoding in the near-term because: a) it is tailored to solve certain families of optimization problems to which FEC decoding tasks can be mapped---for instance, mapping the FEC decoding solution to the lowest energy configuration in a high-dimensional energy landscape is shown in Fig.~\ref{fig:intro_1}, and b) it is reasonably robust to qubit noise, a must for near-term utility.
While a naive transformation of the FEC decoding problem to an optimization form amenable to QAOA is, in theory, sufficient to solve the problem, this is insufficient in practical wireless network settings due to the following issues: First, the iterative nature of the QAOA algorithm (see \S\ref{bkg_VQA}) imposes large problem processing latency by requiring numerous iterations for convergence to solution, and second, quantum operations (gates) are currently much slower than their classical counterparts. These issues make the real-time use of QAOA very challenging, particularly for time-sensitive applications with strict deadlines such as FEC decoding (\textit{ca.} 1--50~$\mu$s is available for FEC decoding in 5G).

To address these issues and to enable efficient decoding, this paper presents \sysname{}, a full-stack QAOA-based decoder design for the popular wireless Low Density Parity Check (LDPC) and Polar FEC codes. \sysname{} accelerates traditional QAOA optimization process by exploiting \textit{temporal similarity} among the FEC decoding problems. In particular, we identify that as wireless data frames arrive at the receiver over time, their corresponding FEC decoding problems bring substantial similarity in their optimization forms due to the fixed structure of the particular employed FEC code. For an FEC decoding task on $N$-variables, distinct problems differ only in $O(N)$ parameters among a total of $O(N^2)$ parameters that characterize the optimization task. This results in highly similar optimization contours among the decoding problems, irrespective of the noise and interference present in the wireless channel. \sysname{} leverages this temporal similarity in determining good QAOA initialization parameters `offline'--performing a one-time extensive QAOA search on a small set of problems available in the wireless packet preamble whose solutions are known a priori. The same initialization parameters are then used to \textit{warm-start} real-time QAOA for decoding the actual FEC payload data, thus accelerating the QAOA process.

We have evaluated \sysname{} for small-scale LDPC and Polar FEC codes with short block lengths of eight bits, in Gaussian wireless channels, showing that \sysname{} on average reduces up to 2.5--3$\times$ of the decoding bit errors in comparison to traditional QAOA methods, while achieving at par error performance in comparison to state-of-the-art classical FEC decoders. We also evaluate \sysname{} in a `one-iteration real-time deployment' scenario, which provides near-optimal results in just one QAOA iteration thanks to the highly accurate warm-starts. Our further studies present a holistic resource estimation analysis, quantitatively determining the required qubit count and quantum operation (gate) durations, for the application of \sysname{} in practical wireless networks, highlighting the scenarios where \sysname{} may outperform classical decoders. Our studies may be of potential interest to NextG quantum systems and NextG wireless networks.

%% file: 2_primer_coding.tex
\section{Primer: Wireless Channel Coding}\label{sec:primer_coding}
\begin{figure}[t]
    \centering 
    \includegraphics[width=0.75\columnwidth]{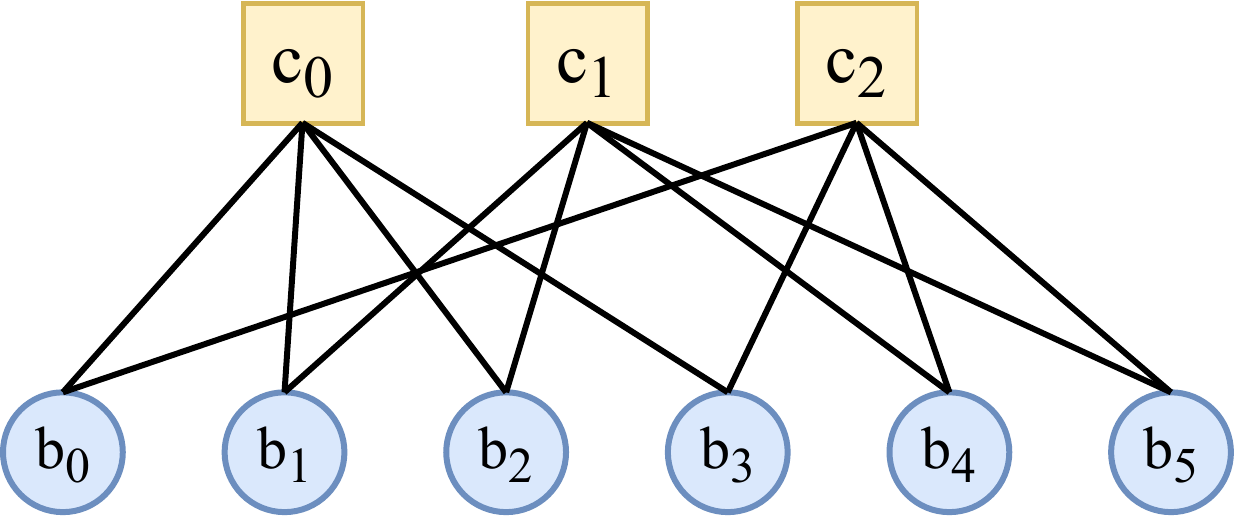}
    \caption{Tanner graph representation of an example LDPC code. Nodes labeled $b_i$ and $c_i$ are bit nodes and check nodes respectively.}
    \label{fig:ldpc_enc}
\end{figure}

This section provides background on FEC codes in wireless networks: Low Density Parity Check (LDPC) codes (\S\ref{s: ldpc}) and Polar codes (\S\ref{s: polar}), describing their encoding and decoding schemes, followed by their classical decoding limitations.

\subsection{LDPC Codes}
\label{s: ldpc}

A binary LDPC code is a linear block code described by a sparse parity check matrix \textbf{H} = $[h_{ij}]_{M\times N}$, where rows and columns in \textbf{H} define \textit{check nodes} and \textit{bit nodes} of the code respectively \cite{gallager1962low}. \textbf{H} can be equivalently visualized as a bipartite graph (called Tanner graph \cite{tanner1981recursive}) shown in Fig.~\ref{fig:ldpc_enc}, where nodes labeled $b_i$ and $c_i$ are bit and check nodes respectively, and an entry $h_{ij} = 1$ in \textbf{H} indicates an edge between $c_i$ and $b_j$ in the Tanner graph. The encoding process is described next. 

\paragraph{LDPC Encoder} Let \textbf{u} be the user data of length $K$ bits, and \textbf{x} be encoded codeword of length $N$ bits. To perform encoding, first convert \textbf{H} into an augmented form $[\textbf{A}|\textbf{I}_{N-K}]$ through elementary row operations, where \textbf{A} is obtained in the conversion process and \textbf{I} is the identity matrix \cite{nguyen2019efficient}. Next construct the generator matrix \textbf{G} = $[\textbf{I}_{K}|\textbf{A}^T]$. The encoded codeword is then \textbf{x} = \textbf{uG}. This way of encoding results in zero checksum (\textit{i.e.,} modulo two sum of bits at every check node is zero) \cite{gallager1962low}.

\paragraph{LDPC Decoder} LDPC codes are traditionally decoded via the \textit{Belief Propagation} (BP) algorithm, which works by iteratively exchanging information between check and bit nodes of the code (see Ref.\cite{hailes2015survey}). The decoding process begins by initializing all the bit nodes with their respective \textit{a priori} log-likelihood ratios (LLRs) computed from the received wireless data. Bit nodes pass this LLR information to the check nodes they are connected to in the Tanner graph. The following two steps are then iterated: \textbf{(1)} All the check nodes compute LLRs to send w.r.t each bit node they are connected to by enforcing the zero checksum condition. \textbf{(2)} All the bit nodes then update their LLR values to send back w.r.t each check node they are connected to by adding the LLRs received from check nodes to its a priori LLR. The decoding terminates at a threshold number of iterations. At the end of decoding, bit nodes make either a zero-decision or a one-decision based on the sign of their LLR value (see Ref.\cite{hailes2015survey}).

\paragraph{Limitations of BP decoder} The drawbacks of the BP decoding approach arise from its FPGA/ASIC hardware implementation, which we summarize here. To achieve accurate decoding and high decoding throughput, high LLR bit precision and high decoding parallelism is required in the hardware respectively. This demands significant amount of hardware resources with complicated resource routing requirements, which together restrict the decoder operational clock speed and hence the decoding throughput \cite{kasi2020towards}. The iterative nature of the BP algorithm limits throughput by requiring numerous serial iterations for convergence--there exist problems that do not converge even for 1000 iterations \cite{bian2014discrete}. These issues therefore compromise decoder throughput and accuracy with hardware implementation \cite{hailes2015survey}. Due to these issues, most practical LDPC decoders are implemented in partly-parallel architectures with low LLR bit precision, thus sacrificing performance\cite{kasi2020towards, hailes2015survey}.

\subsection{Polar Codes}
\label{s: polar}

\begin{figure}
    \centering \includegraphics[width=0.6\columnwidth]{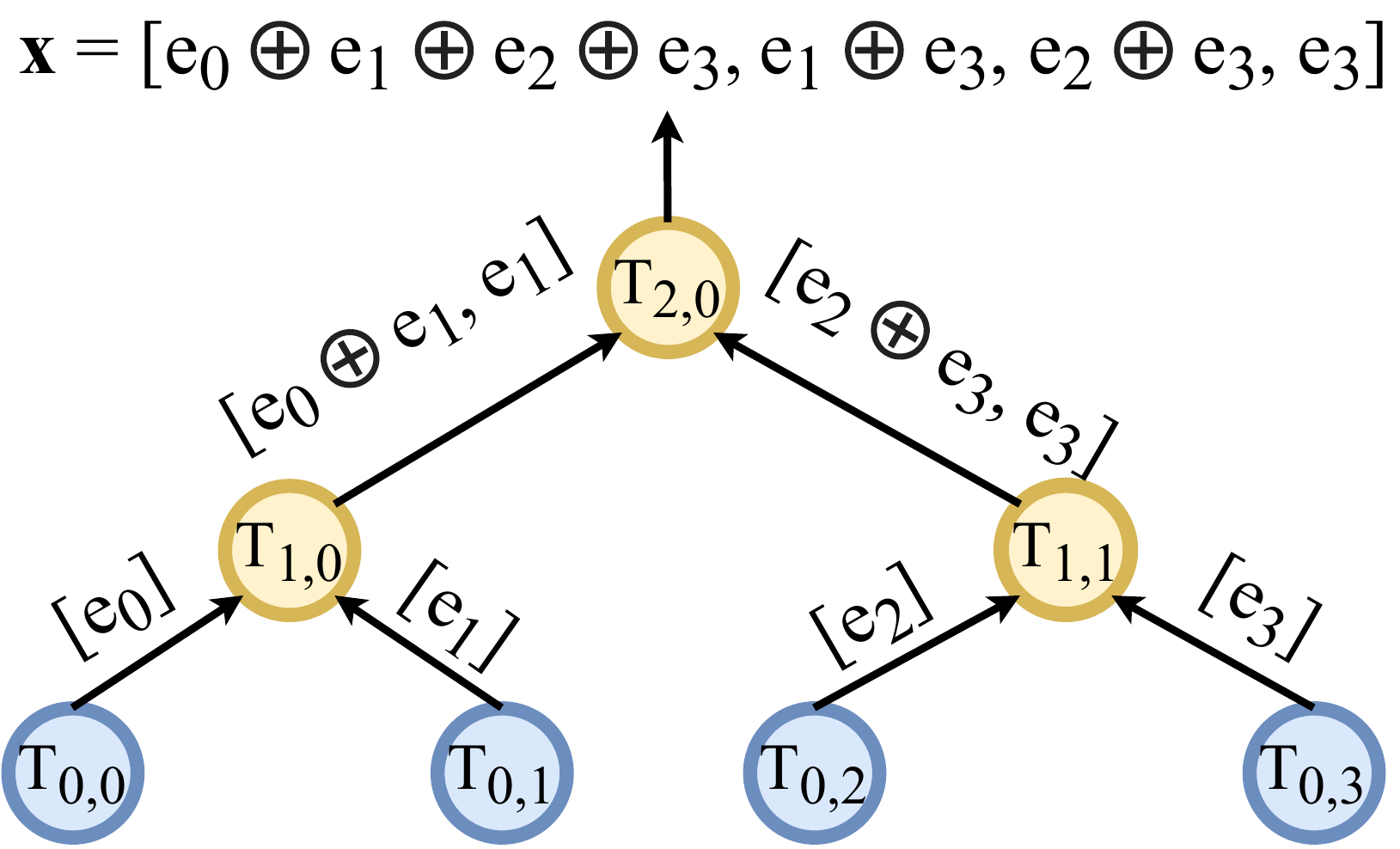}
    \caption{Encoding process of an example 4-bit Polar code. $T_{i, j}$ is a node of the tree at height $i$ and index $j$. Figure adapted from Ref.\cite{10246149}.}
    \label{fig:polar_enc}
\end{figure}

A binary Polar code is a linear block code described by a generator matrix $\textbf{G}_N = \textbf{G}_2^{\otimes d}$, where $N$ = $2^d$ is the code's block length, $\textbf{G}_2 = \big[\begin{smallmatrix}
  1 & 0\\
  1 & 1
\end{smallmatrix}\big]$, and $\otimes d$ is \textit{d}-fold Kronecker product \cite{arikan2009channel}. The overall encoding process can be visualized in Fig.~\ref{fig:polar_enc} which we describe next.

\paragraph{Polar Encoder} Let \textbf{u} be the user data of length $K$ bits. To perform encoding, the encoder constructs an $N$ bit input vector \textbf{e}, where bits in \textbf{u} are assigned to \textit{K} most reliable locations in \textbf{e} and the remaining $N-K$ bits in \textbf{e} are assigned to zero value.\footnote{ Such reliable locations are standardized and this work adapts the standardization of 5G-NR \cite{3gpp}.} The bits that are assigned to zero-value are called \textit{frozen bits}. The encoded codeword is then \textbf{x} = \textbf{e}$\textbf{G}_N$.

To understand the encoding process, let us consider an example Polar code with input vector \textbf{e} = $[e_0, e_1, e_2, e_3]$. Figure~\ref{fig:polar_enc} summarizes the \textbf{x} = \textbf{e}$\textbf{G}_4$ computation. Build a perfect binary tree, initializing leaf node $T_{0, i}$ with bit $e_i$. Next traverse the tree bottom--up, where each node $T_{i, j}$ of the tree transforms the input $[e_L | e_R]$ to the output $[e_L\oplus e_R, e_R]$. Here, $e_L$ and $e_R$ are the output vectors of the left and right children of $T_{i, j}$ respectively, and $\oplus$ is a bit-wise XOR operation. The output vector obtained at the root node is the resulting encoded codeword \textbf{x} (see Fig.~\ref{fig:polar_enc}) \cite{10246149, bioglio2020design}.

\paragraph{Polar Decoder} Polar codes are traditionally decoded via the \textit{Successive Cancellation List} (SCL) decoder \cite{tal2015list}, which works in a depth-first search fashion and decodes one bit at a time. The decoding process begins at the root node of the code's binary tree by initializing the \textit{a priori} LLR values of the root node bits computed from the received wireless data. Each node in the tree then sends to its children the LLR values of their corresponding bits, and the tree traversal follows depth-first with priority given to the left branches. In this process we obtain the LLRs of the leaf node bits sequentially. If a leaf node corresponds to a frozen bit, a zero decision is made. Otherwise, the decoder expands into two paths, one for zero-decision and one for one-decision. The number of decoding paths therefore grows exponentially with $K$. To reduce computational complexity, only best $L$ paths are maintained throughout the decoding ($L$ is called the list size), rest are pruned. At the end of decoding, the path with high likelihood of being correct is selected as the decoded answer (see Ref.\cite{tal2015list}).

\paragraph{Limitations of SCL decoder} SCL decoders by nature inherently sacrifice performance by pruning the decoding paths. For a Polar code with $K$ user data bits, optimal decoding performance is obtained when the SCL decoder's list size is $2^K$. This is challenging to implement on traditional FPGA/ASIC hardware due to exponential complexity when $K$ is large. Further, the SCL decoding process is sequential which means that its latency increases at least linearly with code block length. Due to this decoding difficulty, Polar code use in 5G-NR is currently limited to control channels with short block lengths ($2^5$--$2^{10}$ bits) \cite{3gpp}. Improving the decoding latency while maintaining accurate error performance would make Polar codes viable candidates for high speed data channels in NextG wireless networks \cite{kasi2022design}.

\begin{figure}
    \centering \includegraphics[width=0.9\columnwidth]{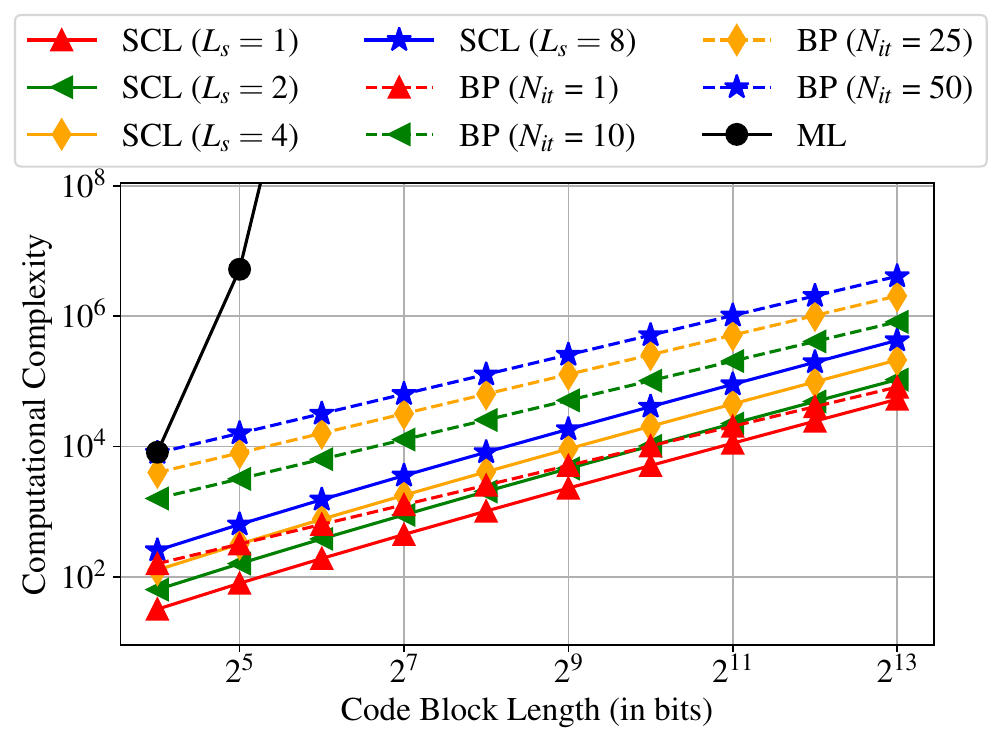}
    \caption{Computational complexity (arithmetic operation count) of various FEC decoders. $L_s$ is SCL decoder's list size, and $N_{it}$ is BP decoder's iteration count. ML decoder provides optimal performance.}
    \label{fig:comp}
\end{figure}

Figure~\ref{fig:comp} shows the computational complexity of BP and SCL decoders for 0.5 rate codes ($K = N/2$). We see that larger block lengths, which provide lower error rates, entail higher computational complexity. The optimal decoding performance is obtained via the Maximum-Likelihood (ML) decoding, an exhaustive search over all possible solutions. ML decoder's complexity scales exponentially with $K$, thus making it intractable for classical hardware implementation.

%% file: 3_primer_quantum.tex
\section{Primer: Quantum Computing}\label{sec:primer_quantum}

This section provides background on quantum gate computation (\S\ref{s: basics:quantum}) and the QAOA algorithm (\S\ref{bkg_VQA}).

\subsection{Basics of Quantum Gate Computation}
\label{s: basics:quantum}

Quantum computation (QC) involves manipulating qubits using logic gate operations that modify their amplitudes. Unlike classical computation, non-trivial single-qubit operations can be performed with QC, such as $R_x(\theta)$ and $R_z(\theta)$, rotating the state of a qubit around the $x$ and $z$ axes. Multi-qubit interactions are achieved with multi-qubit gates such as the two-qubit Controlled-$X$ or $CX$. Together with single-qubit gates, $CX$ allows for universal quantum computation. 
The current and near-term quantum devices are sensitive to hardware noise and are limited in size in terms of qubit count~\cite{preskill2018quantum}. Some of the biggest challenges that limit QC scalability include limited qubit coherence time, state preparation and measurement (SPAM) errors, gate errors, crosstalk, among others, which all degrade performance~\cite{ravi2022quantum}.
This would mean near-term QC devices will be unable to execute large-scale quantum algorithms that require full fault-tolerance with error correction comprised of millions of qubits~\cite{O_Gorman_2017}.
However, a variety of error mitigation techniques~\cite{czarnik2020error,Rosenberg2021,barron2020measurement,botelho2021error,wang2021error,takagi2021fundamental,temme2017error,li2017efficient,giurgica2020digital,ding2020systematic,smith2021error} are underway, encouraging QC devices to the verge of quantum utility~\cite{kim2023evidence}.

\subsection{QAOA for optimization tasks}
\label{bkg:qaoa}
\label{bkg_VQA}

Quantum Approximate Optimization Algorithm (QAOA) is a quantum algorithm that aims to find the lowest energy configurations of combinatorial optimization problems \cite{farhi2014quantum}. One such class of problems we focus in this work are the Quadratic Unconstrained Binary Optimization (QUBO) problems, specified by the cost function:

\begin{equation}\label{eq:qubo}
    c(x) = x^{T} \textbf{Q} x,
\end{equation}

where $x$ is a length $n$ binary vector (bitstring), and $\textbf{Q}$ is a real valued, symmetric $n \times n$ dimensional matrix. The goal of the optimization task is to find $x$ that minimizes the cost function $c(x)$. To understand a QUBO function in the language of QC: we define a Hamiltonian $C$ such that $C\ket{x} = c(x)\ket{x}$ for all the $2^n$ length $n$ bitstrings $x$. The goal of the quantum algorithm is then to boost quantum amplitudes corresponding to bitstrings $x$, at or near the optimal solution so that sampling the state at the end of running the algorithm can yield optimal or near-optimal results.

Variational quantum algorithms (VQAs) are a popular class of algorithms that leverage quantum and classical computation in a hybrid workflow to solve optimization problems \cite{cerezo21_vqa}. The short quantum circuits in VQAs make them relatively robust to qubit noise and therefore attractive in the NISQ regime \cite{peruzzo2014variational, farhi2014quantum}.
VQAs have potential applications in approximation \cite{moll2018quantum}, optimization \cite{farhi2014quantum}, chemistry \cite{peruzzo2014variational}, among others \cite{biamonte2017quantum}.

QAOA is a widely studied VQA originally introduced for finding approximate solutions to constraint satisfaction problems (CSPs), which include QUBOs. QAOA broadly consists of a circuit (called an \textit{ansatz}) with $p$ alternating parameterized layers acting on some specified initial state $\ket{\Psi_0}$. For each layer $\ell$, the phasing operator $e^{-i \gamma_{\ell}C}$ applies a complex phase to each bitstring $x$ in the quantum state corresponding to its cost $c(x)$ given by the QUBO at hand (Eq.~\ref{eq:qubo}) and some real-valued parameter $\gamma_{\ell}$. This phasing step can be implemented for any QUBO \cite{glover19}. The mixing layer $e^{-i \beta_{\ell}B}$ then combines amplitudes of quantum states with complex coefficients depending on Hamming distance between the states and some real-valued parameter $\beta_{\ell}$. Concretely, $B$ is taken as $\sum_i X_i$, so $e^{-i \beta_{\ell}B}$ can be implemented as single-qubit $R_x$ rotation on every qubit. The entire QAOA circuit is then characterized by \cite{farhi2014quantum}: 

\begin{figure}[t]
    \centering  
    \includegraphics[width=\columnwidth]{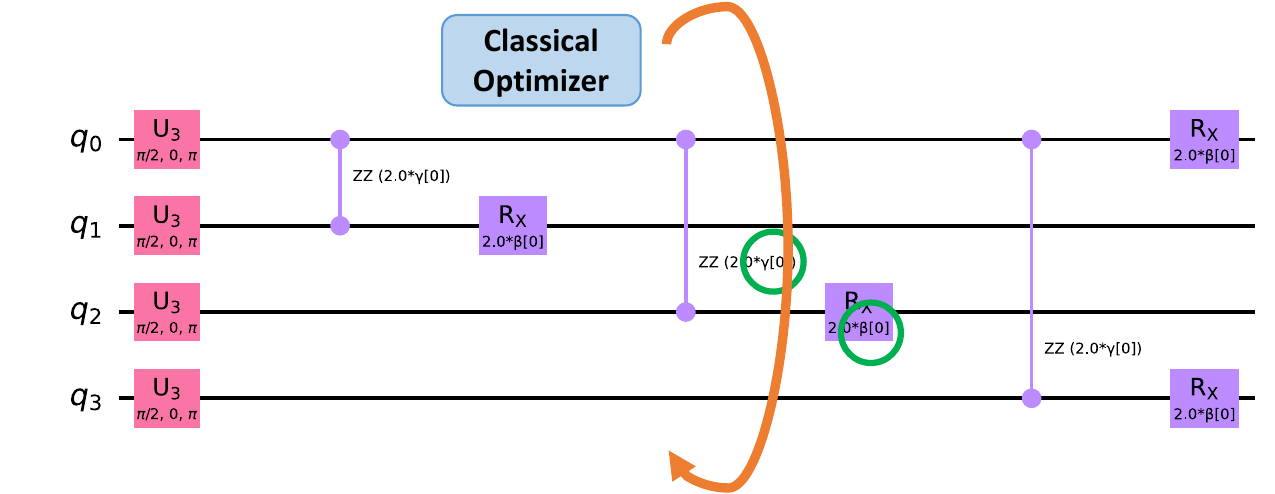}
    \caption{QAOA ansatz with one layer for a 4-qubit graph problem.}
    \label{fig:bkg_1}
\end{figure}

\begin{equation}\label{eq:qaoa_unitaries}
    \prod_{\ell=1}^p \left( e^{-i \beta_{\ell}B} e^{-i \gamma_{\ell}C} \right) \ket{\Psi_0}.
\end{equation}

An example quantum circuit for implementing Eq.~\eqref{eq:qaoa_unitaries} is shown in Fig.~\ref{fig:bkg_1} for a 4-qubit problem.
One must then identify $2p$ parameters $(\vec{\gamma}, \vec{\beta})$ that yields a quantum state in Eq.~\eqref{eq:qaoa_unitaries} with large quantum amplitudes corresponding to optimal or near-optimal bitstrings $x$. 
In the limit of $p\rightarrow \infty$, QAOA converges to quantum adiabatic computation, which can solve any CSP exactly. For small and finite values of $p$ on various CSPs and limits, results have been derived \cite{wang18, farhi22, barak15, hastings19, marwaha21, chou22, lin16, farhi15, hadfield19}.

The performance of QAOA relies on the choice of parameters $(\vec{\gamma}, \vec{\beta})$, and techniques for parameter optimization are in many cases highly non-trivial~\cite{wang18, farhi22, barak15, hastings19, marwaha21, chou22, lin16, farhi15, hadfield19, streif20}. A common strategy is to resort to variational optimization of parameters. In this paradigm, given parameters $(\vec{\gamma}, \vec{\beta})$, a quantum computer is used to estimate the expectation value of the quantum state induced by QAOA with parameters $(\vec{\gamma}, \vec{\beta})$ with respect to the cost operator $C$. A classical computer then uses this cost estimate, in conjunction with previous estimates, to update $(\vec{\gamma}, \vec{\beta})$ for future calls to the quantum computer. This quantum-classical loop is performed until some convergence criteria is met. In practical applications employing VQAs, lowering the time to accurate convergence (\textit{i.e.,} low execution time) is vital for quantum based techniques to compete and eventually outperform state-of-the-art classical methods~\cite{cafqa}.

One technique is based on the principle that globally optimal parameters for QAOA instances randomly drawn from the same distribution tend to concentrate \cite{brandao18, farhi2014quantum}. Thus, if globally optimal parameters are known for one, or many instances of a specific class of problems, these  parameters empirically perform well on a new, unseen instance from the same class \cite{galda21, shaydulin21, sud22}. Our work takes inspiration from this observation. In Section~\ref{s: systemdesign}, we note that our target FEC decoding problems have the exact same problem structure with a low coefficient variance among distinct problems, and so these similarities allow for previously-found ansatz parameters to transfer even more successfully to new problem instances. We refer to the QAOA algorithm described this section as \textit{traditional QAOA} in the rest of the paper.

%% file: 4_design.tex
\section{Design}\label{sec:design}

We now describe the QUBO forms of LDPC and Polar FEC codes (\S\ref{design:qubo}) \cite{kasi2020towards, kasi2022design, bian2014discrete, chancellor2016direct}, and then present our \sysname{} system design (\S\ref{s: systemdesign}), alongside our practical considerations (\S\ref{s: parameters}).

\subsection{Wireless FEC decoding as a QUBO problem}
\label{design:qubo}

Recent prior work in this area has formulated the LDPC and Polar decoding problems into QUBO forms, which we briefly summarize here \cite{kasi2020towards, kasi2022design, bian2014discrete, chancellor2016direct}.
To obtain QUBO forms, the core idea is to construct two types of cost penalty functions: a \textit{satisfier} function to characterize the structural computation of the particular code (\textit{c.f.,} Figs.~\ref{fig:ldpc_enc}, \ref{fig:polar_enc}), and a \textit{distance} function to capture the proximity of a candidate bit string to the received wireless data. The entire QUBO is then a weighted linear combination of these cost penalty functions. The bit string with minimum cost penalty is the decoded solution.

For both the FEC codes, the satisfier function value is minimum for all the candidate bit strings that agree with the respective encoding conditions. This agreement corresponds to satisfying the zero checksum condition in LDPC codes (\S\ref{s: ldpc}), and satisfying the binary tree computation in Polar codes (\S\ref{s: polar}). If a candidate bit string disagrees, a cost penalty is raised for that candidate. Therefore, the satisfier function eliminates invalid bit strings (\textit{i.e.,} non-codewords) as potential solutions to the problem. The distance function is minimum for the candidate bit string that is most proximal to the received wireless data, thus it prefers the most-likely transmitted bit string to be the decoded solution. The entire QUBO objective function value is therefore minimum for the codeword that is closest to the received wireless data, which is essentially the Maximum-Likelihood solution to the problem. We refer the reader to Refs.\cite{kasi2020towards, kasi2022design, bian2014discrete, chancellor2016direct} for further details.

\subsection{\sysname{}: Proposed QAOA-based FEC Decoder}
\label{s: systemdesign}

\begin{figure}
    \centering
    \includegraphics[width=0.85\columnwidth]{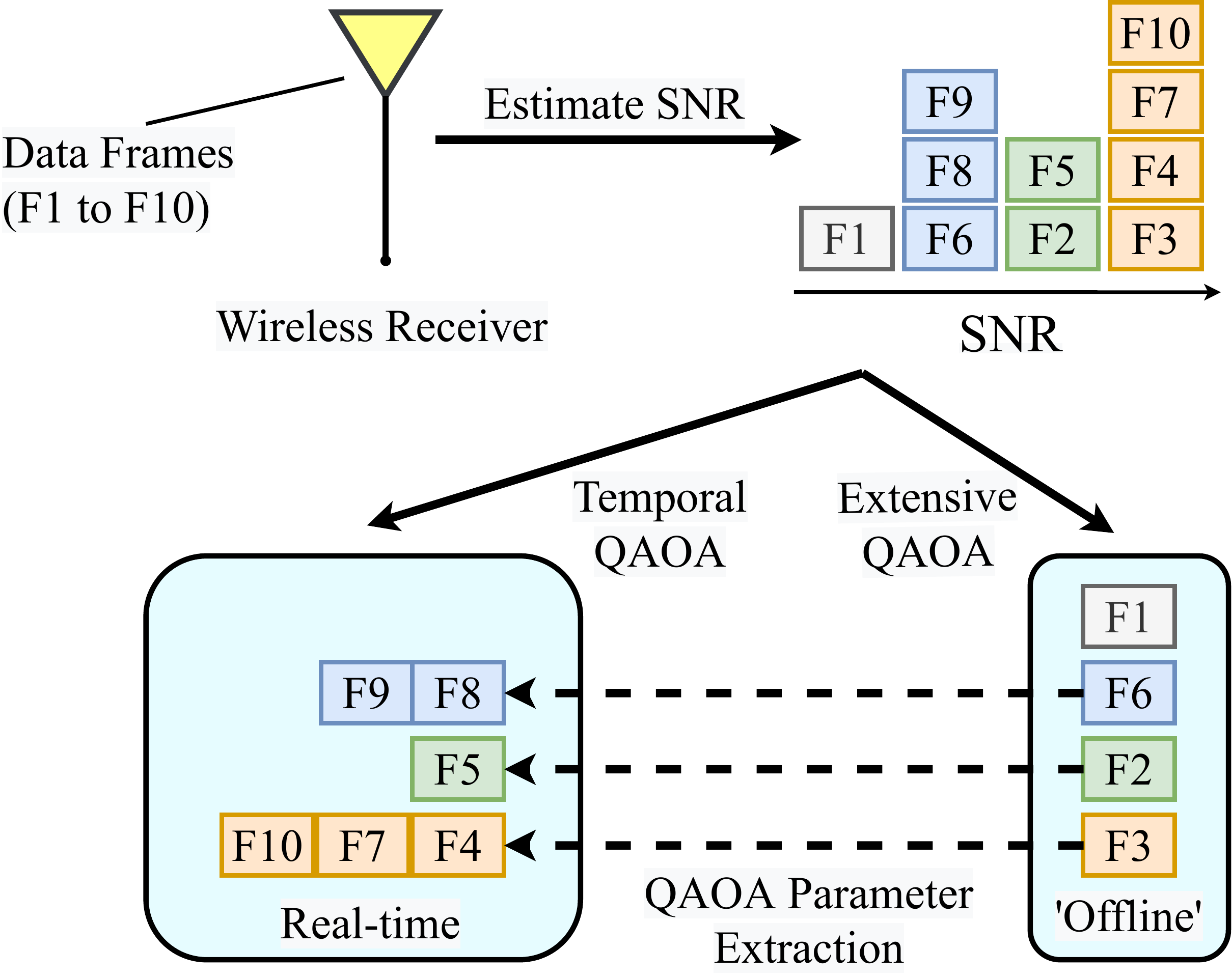}
    \caption{\sysname{}'s decoder architecture. As wireless data frames arrive at the receiver over time, \sysname{} groups them by their SNR. One frame per SNR is decoded via extensive QAOA search `offline', determining their optimal QAOA parameters. Rest of the frames at the same SNR use the same QAOA parameters for real-time decoding.}
    \label{fig:designfig}
\end{figure}
\sysname{} is a full-stack QAOA-based decoder design for LDPC and Polar FEC codes, which are becoming widespread in 5G and NextG wireless networks \cite{3gpp}. The end-to-end working process of \sysname{} is depicted in Fig.~\ref{fig:designfig}.

In wireless networks, \textit{data frames} carry user payload data. Each data frame consists of multiple \textit{blocks}, where each block contains a portion of payload data that is encoded via a particular FEC code. In order to decode these blocks, the receiver considers its employed FEC code and the estimated signal-to-noise ratio (SNR) that characterizes the payload data corruption due to noise and interference in the wireless channel. \sysname{}'s decoding works by grouping the data frames by their SNR information as shown in Fig.~\ref{fig:designfig}. At each SNR, we have multiple frames which share significant temporal information (detailed below). \sysname{} exploits this temporal similarity by decoding one frame per SNR via an extensive QAOA search `offline', determining their optimal QAOA initialization parameters, and then re-using the same parameters for decoding the rest of frames within the same SNR interval in real-time. The `offline' search is simplified by utilizing \sysname{} frames whose solutions are known a priori.

\begin{figure}
    \centering
    \includegraphics[width=0.86\columnwidth]{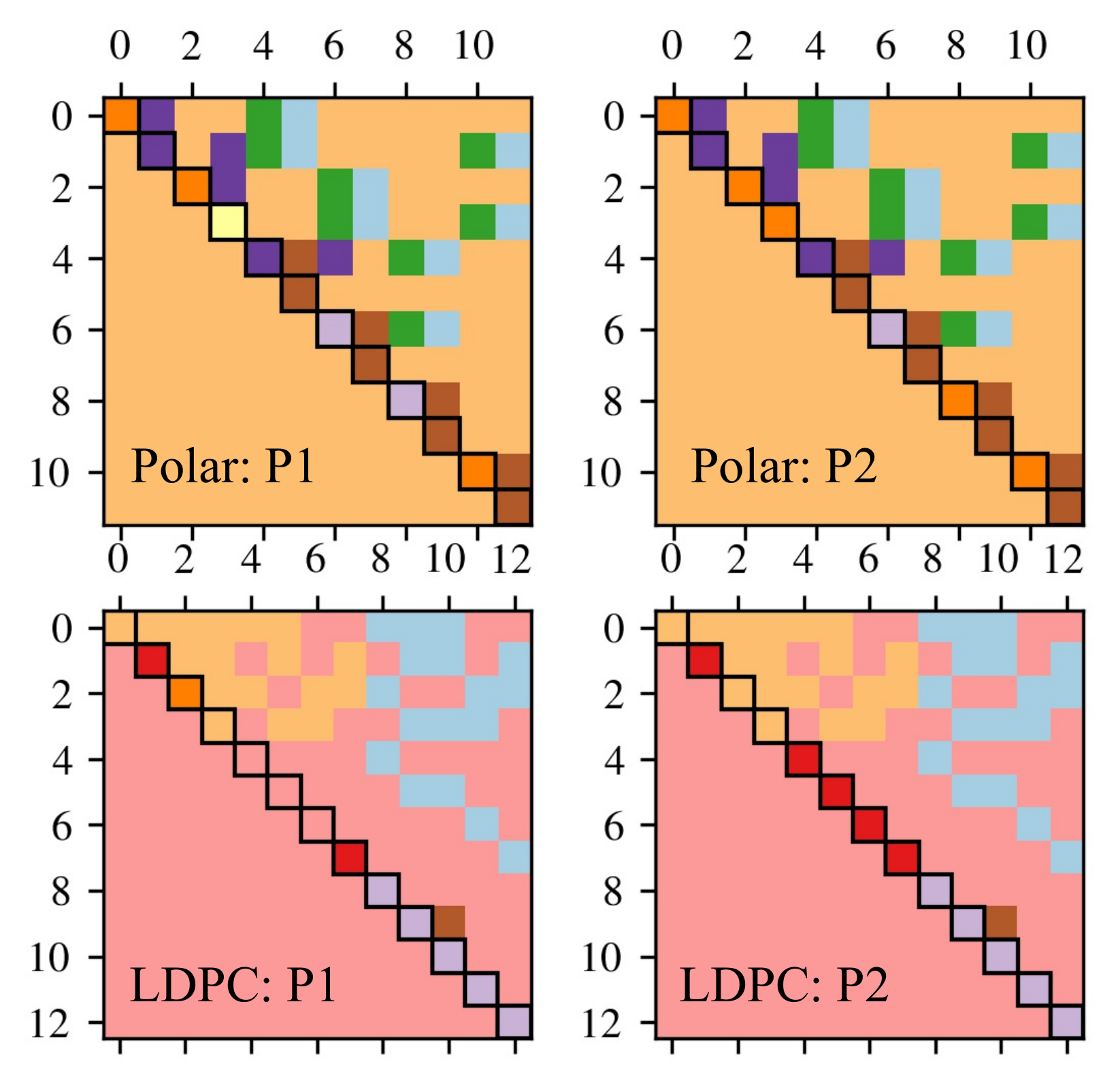}
    \caption{QUBO matrices of LDPC and Polar codes, showing that only a fraction of diagonal entries (black borders) differ from problem (P1) to problem (P2). Similar property holds for any number of problems. Different colors indicate different coefficient values.}
    \label{fig:mats}
\end{figure}

\paragraph{Understanding Temporal Similarity} Figure~\ref{fig:mats} shows the decoding QUBO matrices of LDPC and Polar FEC codes, with real data for two example problems (\textit{i.e.,} blocks) each. We observe that for a given FEC code, distinct problems differ only in a fraction of diagonal entries, while the off-diagonal entries remain the same across problems. A similar property holds for any number of problems. This is because the off-diagonal entries depend only on the particular fixed FEC code, whereas the diagonal entries depend on the time-varying wireless channel noise and interference which is characterized by SNR \cite{kasi2020towards, kasi2022design, bian2014discrete, chancellor2016direct}. Therefore for a given SNR, this results in highly similar optimization contours, indicating the existence of temporal similarity across the equivalent QAOA tasks.

\begin{figure}
    \centering  
    \includegraphics[width=\columnwidth,trim={1cm 0cm 3cm 0cm},clip]{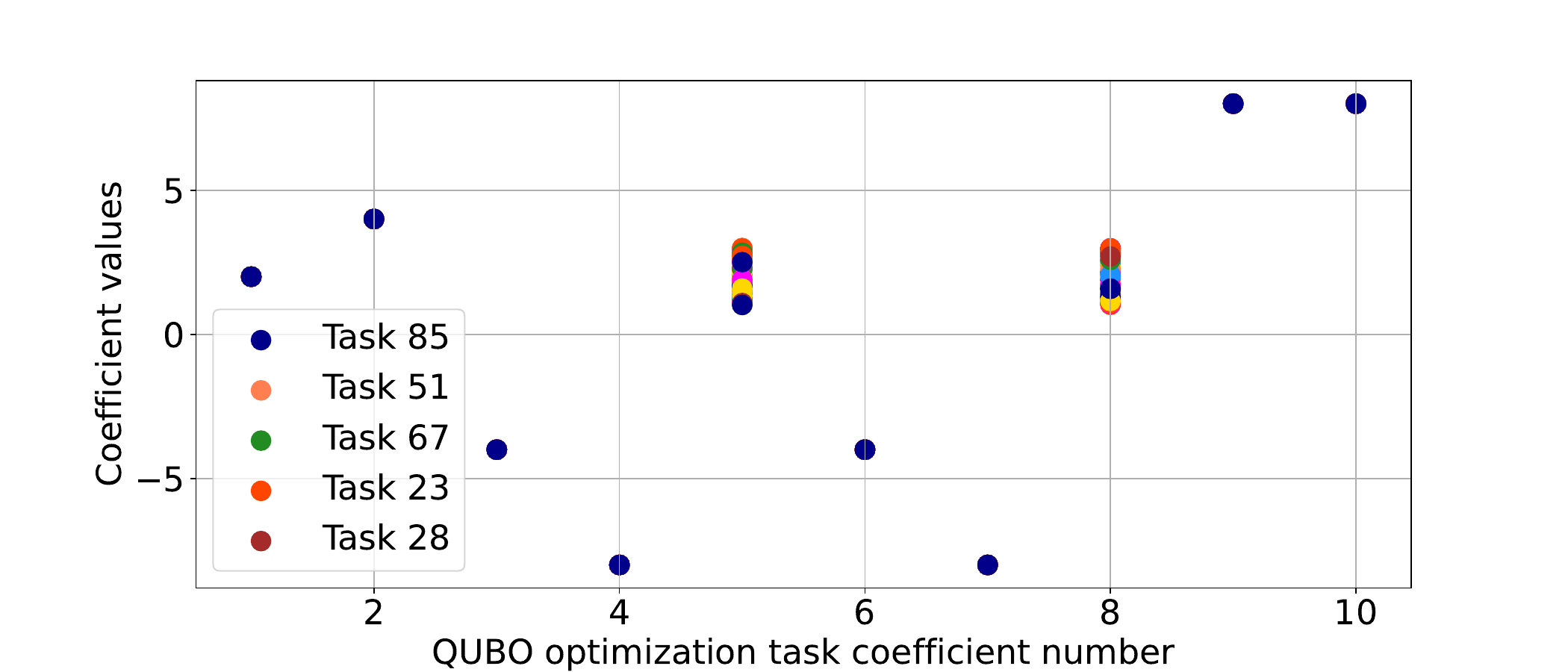}
    \caption{QUBO coefficient index vs. value for 100 4-bit Polar decoding problems. For clarity, only five tasks are labeled. The figure shows eight out of 10 coefficients remain the same across problems while two of them vary. Invisible data points are overlapping.}
    \label{fig:design_c}
\end{figure}

\paragraph{Exploiting Temporal Similarity} \sysname{} exploits temporal similarity via a ``warm-start'' QAOA approach by pre-determining the QAOA initialization parameters (\textit{i.e.,} $\vec{\gamma}, \vec{\beta}$) that are best tailored to the decoding task at hand. In particular, we choose the same QAOA initialization parameters for decoding frames within the same SNR value, thus efficiently exploiting temporal similarity in the decoding process. This approach is highly effective 
for FEC decoding for the following two reasons which was motivated earlier in Fig. \ref{fig:mats}.
First, because the nature of QUBO connectivity remains same across problems (\textit{i.e.,} off-diagonal entries in Fig.~\ref{fig:mats}), the structural contour of the QAOA optimization process remains the same across distinct problems.
Second, and even more critical, is that a significant fraction of QUBO coefficient values remain the same for a particular FEC code, independent of the wireless channel noise, ambient interference, and the data being encoded itself. For a QUBO problem on $N_v$ variables, only $O(N_v)$ coefficients differ from problem to problem among a total of $O(N_v^2)$ coefficients. 
Third, the differing coefficients across problems themselves vary by fairly small magnitudes. 
All of this results in a highly similar QAOA search process among problems.
To illustrate this, Fig. \ref{fig:design_c} depicts QUBO coefficient index on the X-axis and their corresponding values on the Y-axis for 100 four-bit Polar decoding problems. Out of 10 coefficients in this setting, eight of the coefficients remain constant across all tasks while the other two vary by less than 20\% magnitude.

We now describe \sysname{}'s working process. As described earlier, \sysname{} first runs extensive QAOA on offline frames using significant computational resources, and finds the optimal parameters of the QAOA ansatz (\textit{i.e.,} $\vec{\gamma}, \vec{\beta}$). Importantly, this computation is off the real-time critical path for wireless decoding, meaning that the offline QAOA search is performed on packet preamble frames and not the actual user payload data. \sysname{} next initializes the same QAOA ansatz parameters for rest of the real-time decoding problems (\textit{i.e.,} user payload data) and performs a local QAOA optimization. Due to the existence of temporal similarity among problems described above, the optimal QAOA parameters for all the decoding problems tend to concentrate, thus enabling QAOA to find the best ansatz parameters for all the problems in real-time frames. The local optimization can be performed either via gradient descent or more sophisticated local approaches~\cite{9259985}. Our experiments show that even when no further local optimization is performed and the pre-computed ansatz parameters are used as is,  \sysname{} is able to achieve excellent decoding accuracy. We refer to this as ``one-iteration real-time deployment''.

\begin{figure}
    \centering  
    \includegraphics[width=0.9\columnwidth]{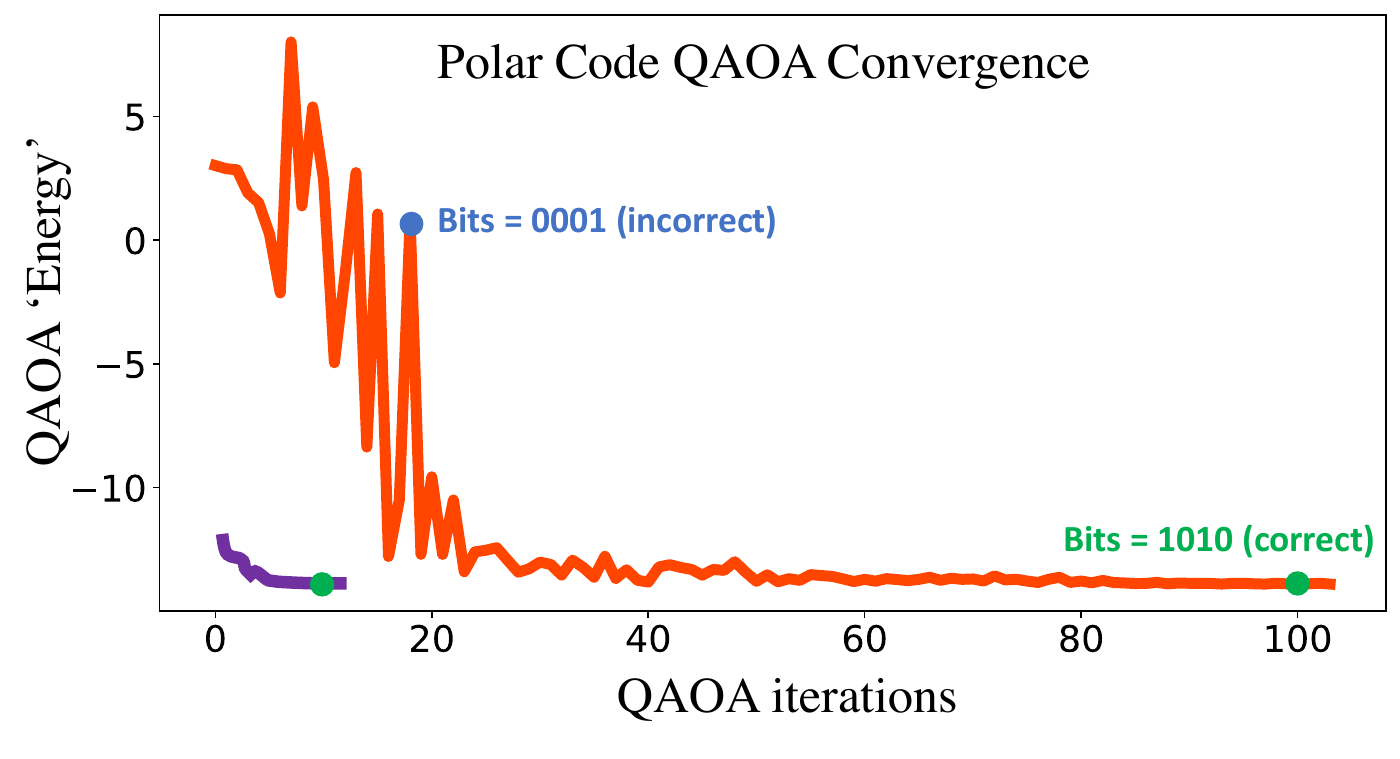}
    \caption{\sysname{} (purple) vs. traditional QAOA (orange) decoding for an example 4-bit Polar code. The figure shows that \sysname{} achieves significantly fast convergence to solution (grey dot) than traditional QAOA. Estimates prior to convergence indicate incorrect decoding.}
    \label{fig:design_B}
\end{figure}

To demonstrate this, Fig.\ref{fig:design_B} reports an example \sysname{} decoding for a 4-bit Polar code, comparing against traditional QAOA with random initialization (\textit{i.e.,} arbitrary $\vec{\gamma}, \vec{\beta}$), in a single layer QAOA ansatz setting. A random initialization implies non-trivial iteration count (and so wall clock time) to convergence. In the figure, we observe that in traditional QAOA the energy value corresponding to the estimated solution fluctuates with high variance prior to convergence, reaching the optimal solution to the problem after a high 100 iterations, whereas \sysname{} achieves the same result in around 10 iterations and each iteration in \sysname{} successively improves upon the candidate solution found in the previous iteration (\textit{i.e.,} decreasing energy from iteration to iteration). For QAOAs (and VQAs) to have a practical utility in the near-term, fast convergence solutions are necessary. While \sysname{} demonstrates this for the FEC decoding problem, we also note that similar ideas may be applicable for generic time-sensitive applications with strict deadlines.

\subsection{Mapping and running on quantum hardware}
\label{s: parameters}

We now provide a hardware perspective of running \sysname{} on current and future quantum devices.
The \sysname{} design is itself independent of hardware but its practical use may be limited by hardware features such as device topology, qubit error rates, and quantum gate duration, which we discuss next.

\textbf{Topology:} Hardware topology refers to the connectivity among qubits. Dense topology (\textit{e.g.,} fully connected trapped ion lattice~\cite{linke17}) allows for quantum circuits to be implemented with lower depth since the logical connectivity required by the algorithm between specific pairs of qubits can be directly implemented on hardware. On the other hand, sparse topology (e.g., heavy-hex superconducting lattice~\cite{ibm21}) limits the direct physical interaction between qubits which require logical connectivity, and therefore requires SWAP operations to physically move the logical qubits closer to one another for interaction. These SWAP operations add additional gates and depth to the quantum circuit, potentially negatively impacting QAOA fidelity via quantum errors. To minimize circuit depth, we adopt state-of-the-art commutativity based QAOA circuit optimizations such as the work from Jin \textit{et al.}~\cite{jin22}, which  achieves low circuit depth for a given hardware topology. Ref.~\cite{jin22} specifically achieves guaranteed linear QAOA circuit depth for typical quantum hardware, for example achieving $6n$ depth for the IBM heavy-hex topology and $5n/2$ for the Google Sycamore topology.

\textbf{Qubit error rates:} While VQAs such as QAOA are somewhat robust to noise, they are still impacted by qubit error rates. Therefore in the near-term we expect to choose the best quality qubits by using typical noise-aware techniques~\cite{murali2019noise}. Further quantum error mitigation techniques such as dynamic decoupling~\cite{pokharel2018demonstration}, measurement error mitigation~\cite{jigsaw}, crosstalk reduction~\cite{murali2020software}, and others tailored to variational algorithms~\cite{ravi2021vaqem, ravi2022navigating, dangwal2023varsaw} will enable efficient \sysname{} decoding.

\textbf{Quantum gate duration:} Gate duration, the execution time of quantum gates, impacts run time of a quantum circuit. In most use cases, the quantum circuit run time primarily matters in the context of decoherence (\textit{i.e.,} qubits losing quantum information over time). However, in real-time use cases, such as FEC decoding, circuit run time considerations are crucial for meeting the 5G/NextG network latency deadlines. In this regard, different quantum hardwares exhibit different gate duration times. For the current status of technology, superconducting and silicon spin qubits have gate operation times in the order of 10s of nanoseconds, photonic qubits can function at around 1 nanosecond, while trapped ion gates are much slower, running into microseconds~\cite{QMod}. Gate durations are typically flexible and can be controlled at the pulse level. We envision the use of prior work on pulse optimizations that aggregates multiple qubit operations into larger pulse blocks that manipulate up to 10 qubits at a time---the effective pulses required for the overall circuit can be shortened by 5--10$\times$~\cite{Shi_2019}. Pulse-level optimization tools such as QuTiP~\cite{Johansson_2012}, JuqBox~\cite{petersson2020discrete}, and Quandary~\cite{günther2021quandary} can be use to achieve optimum trade-offs between gate duration and error rates as suited to particular applications. These tools and pulse-optimization research in general are currently in early stage development and are expected to mature in the near future.

%% file: 5_implementation.tex
\section{Methodology}\label{sec:implementation}

\label{sec:implementation:microbenchmarks}

We conduct \sysname{} system runs via Qiskit \cite{qiskit23} on an Intel i7-10510U CPU@1.80GHz processor with 16GB RAM. Our methodology is to evaluate \sysname{}'s system performance for both LDPC and Polar codes, in the aforementioned QAOA temporal initialization, random initialization (\textit{i.e.,} arbitrary $\vec{\gamma}, \vec{\beta}$), and zero initialization (\textit{i.e.,} $\vec{\gamma} = \vec{\beta} \approx 0$) settings, in both noiseless (exact) and noisy simulations (\S\ref{sec:design}). For the temporal initialization, we optimize the QAOA ansatz parameters for the first QUBO problem within a given SNR by performing local optimization starting from the angles of Ref.\cite{farhi22} with all $\gamma$ values re-scaled by the inverse square root of the sum of QUBO coefficients \cite{shaydulin23}, and then use these parameters to initialize the rest of the QUBO problems. For random initialization, we choose ansatz parameters in the range $[0,\pi]$, and for zero initialization, ansatz parameters are set to $10^{-4}$. After the initialization, parameter optimization is performed using the Constrained Optimization BY Linear Approximation (COBYLA) algorithm function from SciPy \cite{virtanen20} with no restriction on the number of iterations. 
An exception to this is the `one-iteration real-time deployment' scenario (see \S\ref{s: systemdesign}) in which COBYLA iteration count for the real-time QAOA tasks is set to $1$. Each microbenchmark in Section~\S\ref{s: micro} is run for $500$ decoding problems, where we target $13$-qubit LDPC and $12$-qubit Polar decoding problems, which both contain two user data bits. To accelerate the simulation run time, we use Qiskit's matrix product state (MPS) simulator \cite{qiskit23} with a maximum bond dimension set to the number of qubits, which provides approximate solutions to the problem.

We next measure performance of \sysname{} in noisy simulation settings with depolarizing noise. Simulating noisy quantum circuits requires keeping track of density matrices rather than statevectors, squaring the time and space complexity of the simulation. Thus for these experiments, we consider the same specifications as above, but analyze shorter $4$-qubit codes, which contain one user data bit. Further, the `one-iteration real-time deployment' scheme described in \S\ref{s: systemdesign} is used for this evaluation to reduce the number of optimization steps. Our quantum circuits are transpiled to RX, RZ, and CNOT gates, which are native gates for typical quantum hardware. One-qubit and two-qubit depolarizing channels are applied after every RX and CNOT gate, respectively with a fixed error rate. This channel effectively introduces a per-gate error to each gate with a probability given by the error rate. RZ is omitted from errors because for many physical QCs, these gates are implemented in classical software via frame shifts \cite{mckay17}.

%% file: 6_evaluation.tex
\section{Evaluation}\label{sec:evaluation}

This section presents \sysname{}'s system performance evaluation, beginning with microbenchmarks (\S\ref{s: micro}) and `one-iteration real-time deployment' (\S\ref{s: oneshot}). We next present our full-system (\S\ref{s: fullsystem}) and noisy simulation results (\S\ref{s: noisy}).

\subsection{Microbenchmarks}
\label{s: micro}

\begin{figure}
    \includegraphics[clip,width=\columnwidth]{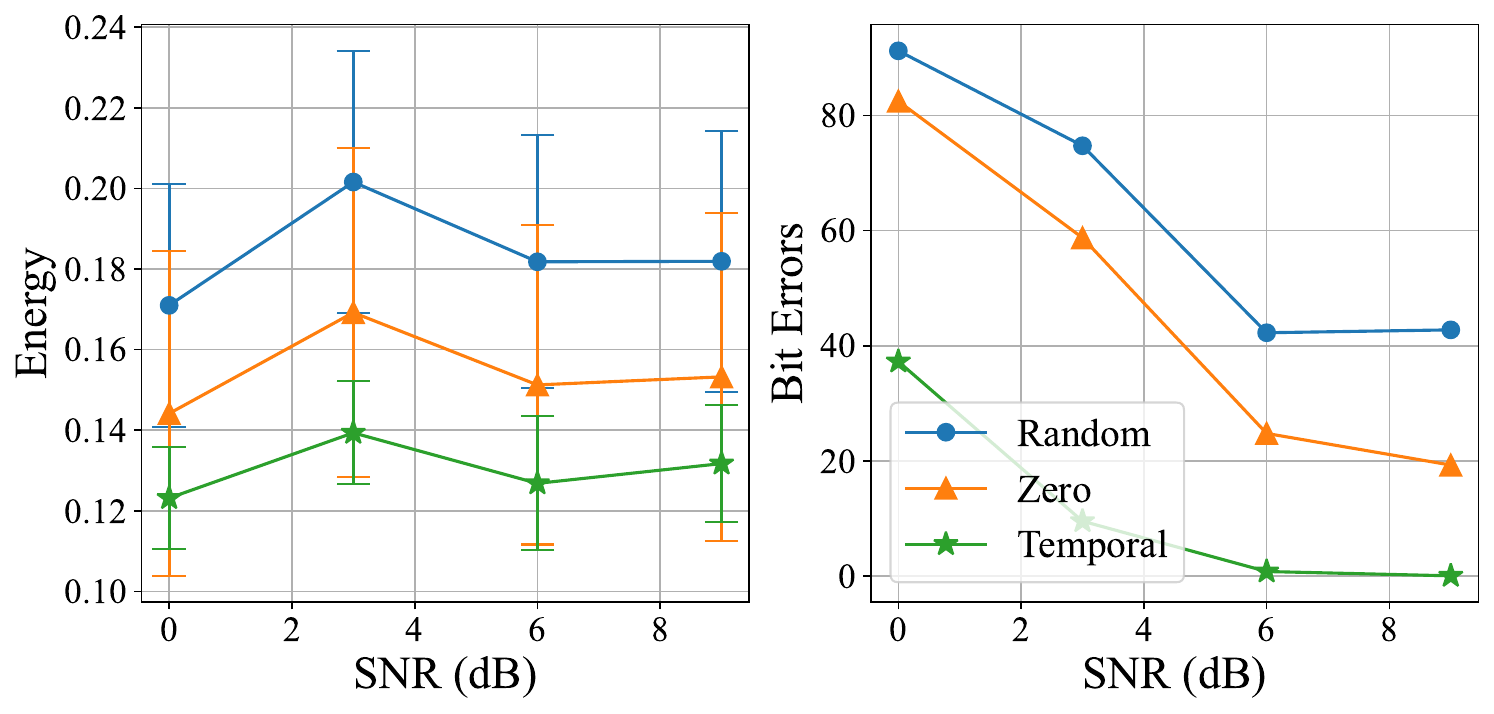}
    \caption{Expected energies (left) and bit errors (right) of solutions returned by QAOA ($p = 4$), for $13$-qubit LDPC codes, at various SNRs and QAOA initializations. Energies are normalized between $0$ and $1$. The figure shows \sysname{}'s temporal initialization outperforms traditional random and zero initialization approaches.}
    \label{fig:ldpc_energies_and_bers}
\end{figure}

Fig.~\ref{fig:ldpc_energies_and_bers} reports the expected energy and bit error counts of \sysname{} under temporal, random, and zero initializations for LDPC codes. The figure (left) shows that at all wireless channel SNRs, temporal initialization provides lower energy solutions and lesser bit errors in comparison to both random and zero initialization methods. The temporal approach reduces energy deviation (from ideal answer) by a median of 20-35\%, and achieves a 3--5$\times$ lower variance in energy compared to zero and random initialization approaches. The magnitude of energy values stay nearly the same across SNRs, indicating that \sysname{}'s temporal initialization provides nearly uniform benefits at all SNRs. Looking at Fig.~\ref{fig:ldpc_energies_and_bers} (right), we observe that \sysname{}'s temporal approach achieves 2.5--3$\times$ lesser bit errors than zero and random initialization methods.

\begin{figure}
    \includegraphics[clip,width=\columnwidth]{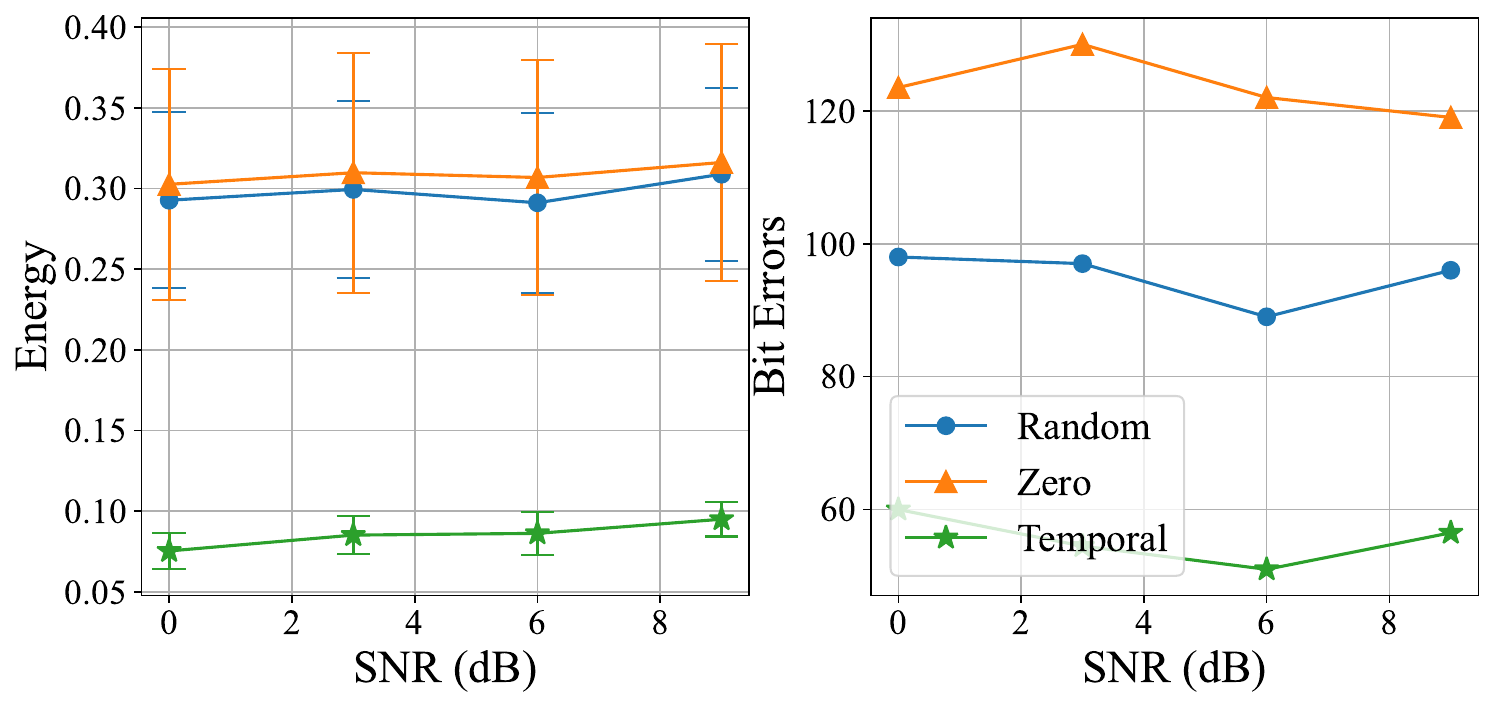}
    \caption{Expected energies (left) and bit errors (right) returned by QAOA ($p = 4$), for $12$-qubit Polar codes, at various SNRs and QAOA initializations. Energies are normalized between $0$ and $1$. The figure shows \sysname{}'s temporal initialization outperforms traditional random and zero initialization approaches.}
    \label{fig:polar_energies_and_bers}
\end{figure}

We next investigate in Fig.~\ref{fig:polar_energies_and_bers} the performance of \sysname{} for Polar codes along the same aforementioned metrics. We observe that similarly as above, \sysname{}'s temporal initialization achieves lower energy solutions and lesser bit errors than random and zero initializations, reducing energy deviation (from ideal answer) by nearly 4$\times$ and variance by 7$\times$. The energy values stay uniform across different SNRs as in LDPC codes (\textit{cf.} Fig.~\ref{fig:ldpc_energies_and_bers}). From Fig.~\ref{fig:polar_energies_and_bers} (right), we note that temporal initialization reduces bit errors by 2--2.5$\times$ compared to random and zero initialization methods.

\subsection{One-iteration real-time deployment}
\label{s: oneshot}

\begin{figure}
    \includegraphics[clip,width=\columnwidth]{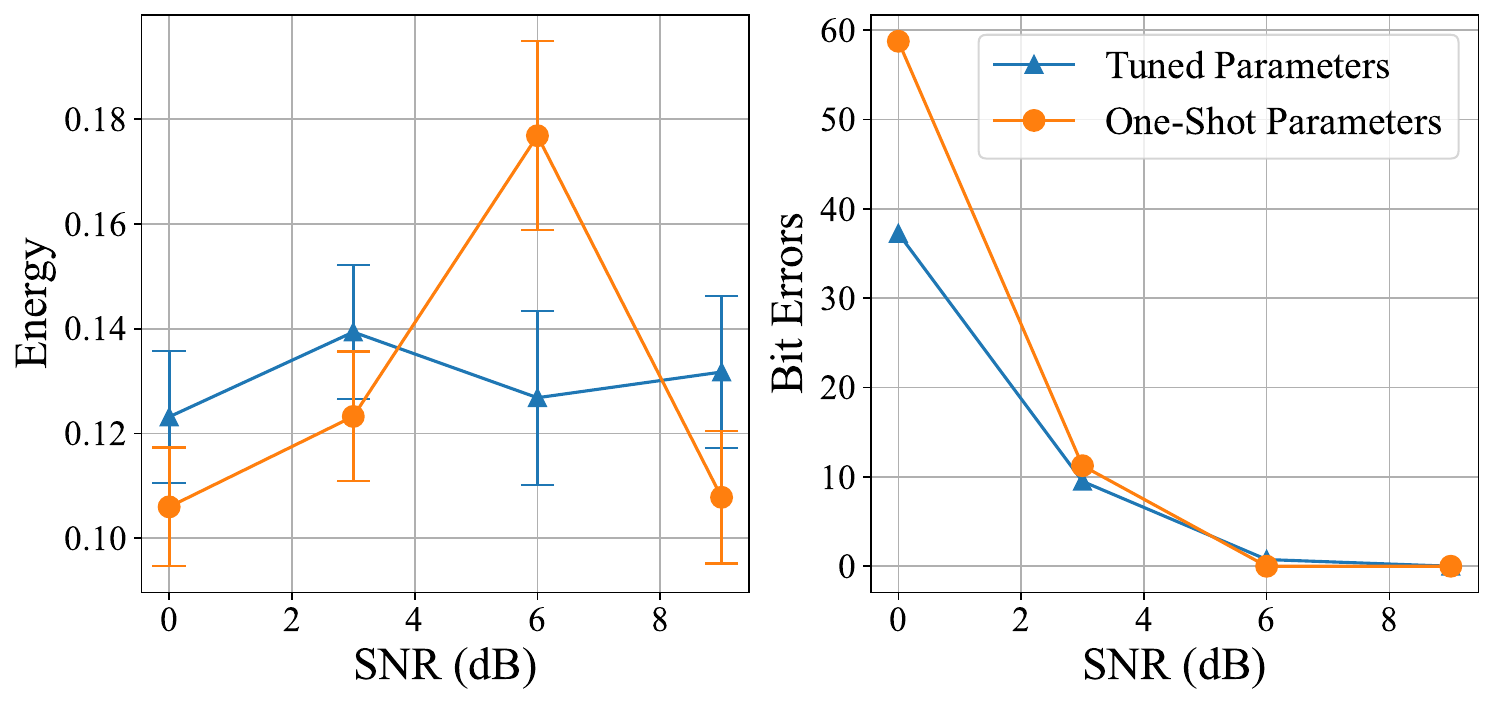}
    \caption{`one-iteration real-time deployment' scenario of Fig.~\ref{fig:ldpc_energies_and_bers}. The figure shows in both one-iteration and tuned-parameter settings with temporal similarity, \sysname{}'s bit error count is barely distinguishable beyond 2~dB SNR. This suggests our one-iteration approach is sufficient to efficiently decode instead of tuning parameters for distinct problems.}
    \label{fig:ldpc_opt_vs_unopt}
\end{figure}

In Fig.~\ref{fig:ldpc_opt_vs_unopt}, we compare \sysname{}'s temporal initialization performance against \sysname{}'s `one-iteration real-time deployment' approach which considers only one QAOA iteration without additional parameter tuning (\S\ref{s: systemdesign}), for LDPC codes along the same metrics described above. The figure shows that both these methods achieve similar solution bit error counts beyond 2~dB SNR, implying that the `one-iteration real-time deployment' approach is promising for practical real-time decoding. Since the `one-iteration real-time deployment' runs only a single QAOA iteration, the problem processing latency is significantly low, thus making it favorable to practical real-time deployment scenarios.

\begin{figure}
    \centering   \includegraphics[width=0.7\columnwidth]{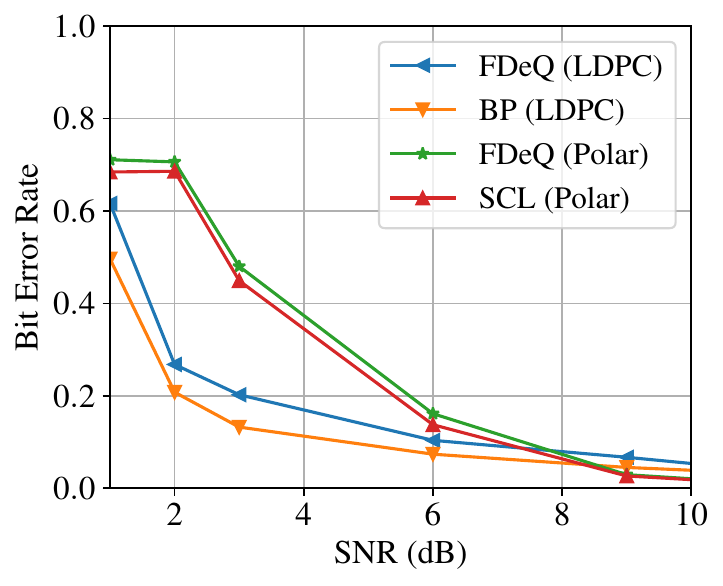}
    \caption{QAOA vs classical decoding error performance}
    \label{fig:polar_qaoa_vs_scl}
\end{figure}

\subsection{Full-system benchmarking against classical decoders}
\label{s: fullsystem}

We next compare the bit error rate performance (the ratio of decoding bit errors to the total number of bits transmitted) of \sysname{} against state of the art classical FEC decoders: BP for LDPC codes and SCL for Polar codes (\S\ref{sec:primer_coding}), referring to Fig.~\ref{fig:polar_qaoa_vs_scl}. Configuration details for the decoders and the experiment can be found in Section \ref{sec:implementation}. The figure shows that \sysname{}'s performance is at par with classical decoders. Simulation restricts us to small block sizes and sub-optimal \sysname{} settings (\S\ref{sec:implementation}), but practical deployment on future quantum devices will allow \sysname{} to scale to larger block sizes employed in practical protocol standards. The associated quantum resource requirements are discussed next in Section \ref{sec:resource_estimation}. On the other hand, classical decoders are not expected to scale well, as shown in Fig.~\ref{fig:comp} which is the primary motivation for this work.

\subsection{Noisy Simulation}
\label{s: noisy}

\begin{figure}
    \centering   \includegraphics[width=0.7\columnwidth]{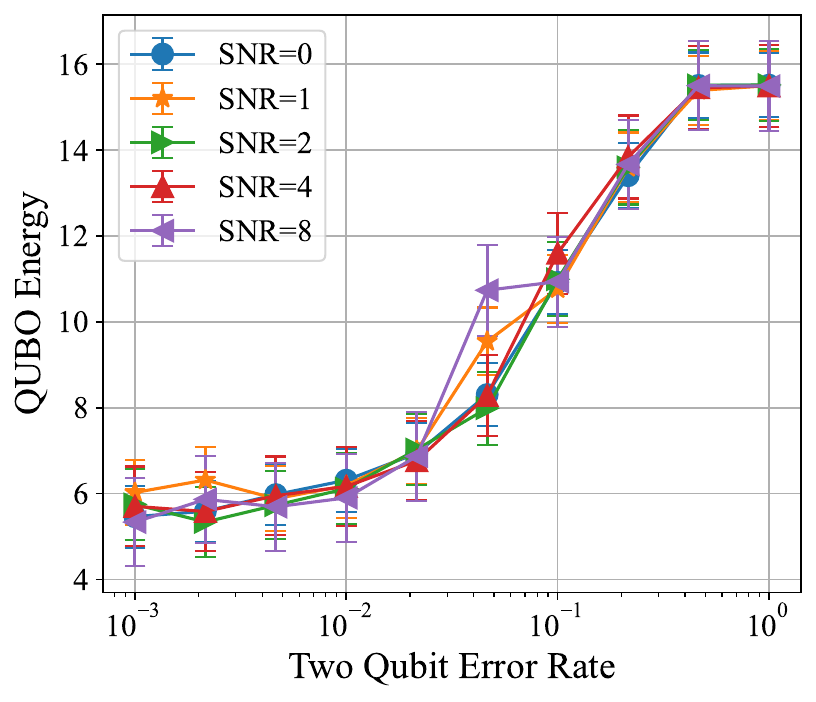}
    \caption{Expected QUBO energy vs SNR vs two-qubit per-gate depolarizing error rate for a $4$ qubit polar code encoding $1$ bit messages. As the error rate decreases between $10^0$ to $10^{-2}$, the energy rapidly converges to the energy given by noiseless simulation (not shown on plot for visual clarity). For many current devices the two qubit error rate is of order $10^{-2}$, suggesting that current-day devices can tackle these small wireless problems.}
    \label{fig:polar_qaoa_snr_vs_error_rate}
\end{figure}

In Fig.~\ref{fig:polar_qaoa_snr_vs_error_rate}, we measure performance of the \sysname{}'s temporal initialization approach in the presence of qubit noise, for a variety of two-qubit gate error rates (X-axis) and SNRs (line trends). From these results, we can see that the QAOA energy estimates are drastically high in the two-qubit gate error rate regime of $10^0$ to $10^{-2}$. Below $10^{-2}$ gate error rate, the energy estimates are much lower, indicating a high correct answer probability. For many current QCs, the average two-qubit error rate is on the order of $10^{-2}$ and approaching to $10^{-3}$ in the near-term. This suggests that near-term QC devices can tackle these small-scale test problems efficiently. As problem sizes scale up, the gate count and circuit depth will increase, so the error rates will need to continue decreasing accordingly to enable QC's performance to not be significantly dominated by noise---note that this is a general requirement for most practical real-world quantum applications.

%% file: 7_resource_estimation.tex
\section{Quantum Resource Estimation}\label{sec:resource_estimation}

This section presents a holistic quantum resource estimation analysis of \sysname{}, projecting quantitative targets for future quantum devices to enable practical FEC decoding, in compliance with the wireless network's timing deadlines.

Let us consider an FEC decoding problem with block length $N$ bits, which is partitioned into $N_{sub}$ number of shorter sub-blocks, where each sub-block of size $N/N_{sub}$ bits is solved via \sysname{}. Note that such partitioning schemes exist for FEC decoding problems and also for generic QUBOs \cite{kasi2022design, glover1977heuristics, qbsolv, glover2010diversification}. Larger sub-blocks provide better decoding performance, and the Maximum-Likelihood performance occurs when $N_{sub} = 1$. Let $N_{it}$ be the number of QAOA iterations, and $N_{ly}$ be the number of QAOA layers per iteration \sysname{} uses to solve each sub-block. The overall run time ($T_{run}$) of a problem is then:
\begin{align}
    T_{run} &= N_{sub}\times T_{run:sub} \\
    T_{run:sub} &= N_{it}\times N_{ly}\times GD \times CD_{ly} \times N_s
\end{align}
where $T_{run:sub}$ is the run time of a sub-block, $GD$ is the two qubit-gate duration, $CD_{ly}$ is the circuit depth of a single QAOA layer, and $N_s$ is the number of circuit shots. $CD_{ly}$ scales proportionally with the number of QUBO variables ($N_v$) present in the sub-block at hand.

\begin{figure}
    \centering   \includegraphics[width=0.9\columnwidth]{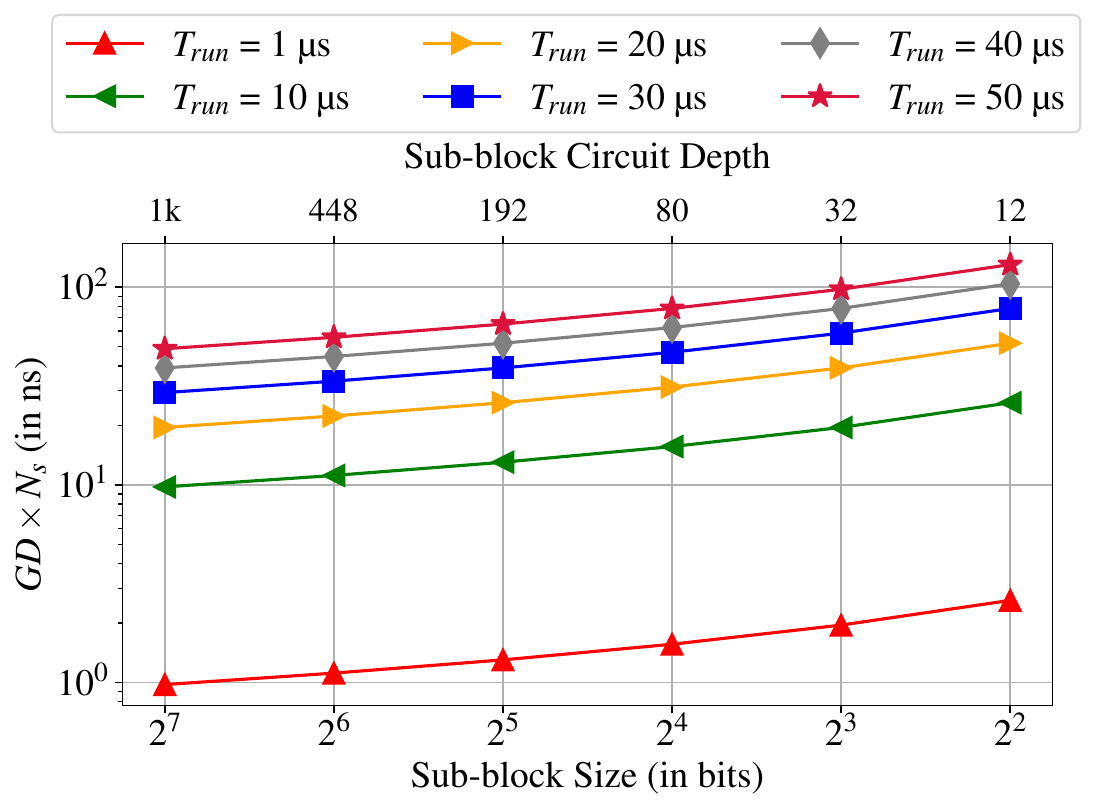}
    \caption{Required two qubit-gate duration to decode 128 bit FEC codes at various sub-block sizes, for \sysname{} to meet the 1--50 $\mu$s decoding deadline.}
    \label{fig:gatelength}
\end{figure}

\paragraph{Gate Duration} The computational time available for decoding a single FEC problem is typically 1--50~$\mu$s in 5G networks. Therefore, to estimate the required gate duration that meets this target, we consider a practical block length of $N = 128$ bits, $T_{run}$ values in [1, 50]~$\mu s$, set fixed values for $N_{it}$, $N_{ly}$ in Eq.~4, and then calculate back GD. Fig.~\ref{fig:gatelength} reports these results with $N_{it} = N_{ly} = 1$. The figure shows that larger sub-blocks require faster gates (\textit{i.e.,} lower two qubit-gate duration) as solving higher complexity problems require deeper QAOA circuits. With 1--130~ns gate duration times and $N_s = 1$, problem run times below 50~$\mu$s can be feasibly achieved for all sub-block sizes. With $N_s = 100$, required gate duration scales down by a factor of 100, indicating that at most 1.3~ns gate duration is needed to meet the 5G latency target. To outperform classical methods via Maximum-Likelihood decoding (\sysname{} achieves this at sub-block size 128 in Fig.~\ref{fig:gatelength}), gate durations should be at most \{48.8, 39, 29.3, 19.5, 9.7, 0.98\} ns when $T_{run}$ requirement is \{50, 40, 30, 20, 10, 1\} $\mu$s respectively.

\begin{figure}
    \centering   \includegraphics[width=0.7\columnwidth]{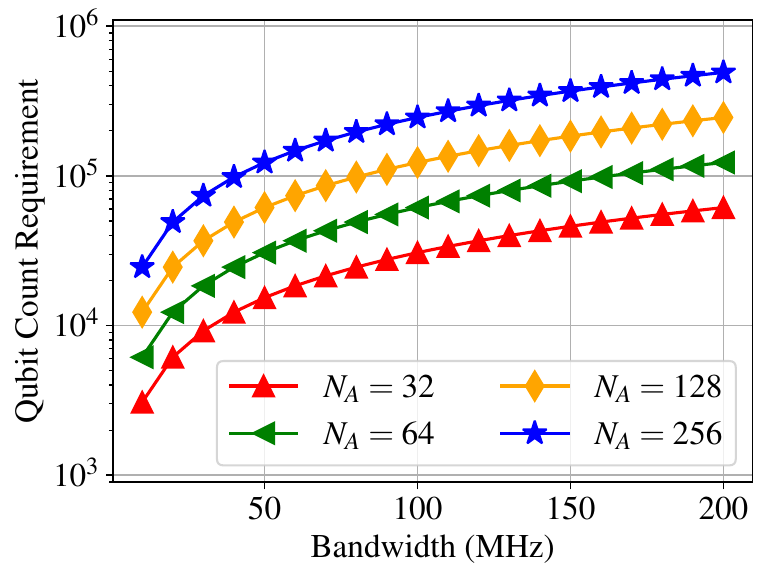}
    \caption{Qubit count requirement for a QAOA processor supporting large MIMO base stations at various bandwidths and antenna count choices, to meet the network's latency and computational demand. $N_A$ is number of antennas.}
    \label{fig:vqa_qubs}
\end{figure}

\paragraph{Qubit Count} While the above gate duration estimates satisfy the network's latency requirement, sufficient amount of qubits are needed in the quantum hardware to satisfy the network's computational requirement. To estimate this required qubit count, we consider the wireless network's FEC computational complexity in estimated \textit{problems per second (PPS)} the BS or C-RAN needs to solve, and problem parameters of a given FEC task. We calculate it as:
\begin{equation}
    N_Q = N_{PPS} \times N_{Q/p} \times T_{run} 
\end{equation}
where $N_Q$ is the number of qubits required for the entire BS or C-RAN FEC processing, $N_{PPS}$ is the number of FEC problems per second the BS or C-RAN needs to solve, $N_{Q/p}$ is the number of qubits required per problem which is esentially the number of QUBO variables, and $T_{run}$ is the run time per problem. For this estimation, we set $T_{run} = 50~\mu s$, which is the typical setting in 5G wireless networks, compute $N_{Q/p}$ by analyzing QUBO forms \cite{kasi2020towards, kasi2022design, bian2014discrete, chancellor2016direct}, and consider PPS values as in Ref.\cite{kasi2021cost}. Figure~\ref{fig:vqa_qubs} reports these results for 10--200~MHz bandwidth base stations at various antenna count choices. To meet the FEC computational demand of near term 5G base stations with 100~MHz bandwidth and 64 antennas, 61k qubits are required in the quantum hardware, whereas the feasibility in processing a small base station with 10~MHz bandwidth and 32 antennas requires only 3k qubits. While these numbers exceed the qubit counts on today's devices, industry roadmaps such as IBM's~\cite{IBM-HW} expect quantum systems with 4k+ qubits by 2025 and 10-100k qubits beyond 2026. Further, our expected requirements are many orders of magnitude lower than the qubit counts needed for fault tolerance~\cite{beverland2022assessing}.

%% file: 8_backend.tex
\section{Conclusions}\label{sec:conclusions}
Forward Error Correction (FEC) in wireless networks ensures reliable communication in the presence of noise and interference in the wireless channel. But it demands exponential computation to achieve optimal decoding performance, which is challenging for classical hardware implementation. Due to such difficulties in the decoder processing requirements, FEC codes employed in practical protocol standards are either limited to short block lengths (\textit{e.g.,} Polar Codes) or their full potential is not realized (\textit{e.g.,} LDPC Codes). Overcoming the limitations surrounding their decoding will enable long FEC codes in NextG communication standards and further enhance the reliability in wireless communication.

With this vision in mind, this paper proposed \sysname{}, a novel QAOA-based decoder design for the popular LDPC and Polar FEC codes. \sysname{} is the first work that leverages temporal similarity via QAOA in order to efficiently decode FEC codes. Our results show that for short FEC codes, \sysname{} outperforms naive QAOA methods with random or zero ansatz parameter initialization, while achieving at par performance in comparison to classical FEC decoders. These results are encouraging, given that quantum computing and simulation are currently at an early-stage development. Our further analysis on a holistic quantum resource estimation shows that quantum operation (gate) duration in 0.1--10 ns range is sufficient to meet the 5G/NextG timing deadline under ideal conditions (\textit{i.e.,} \sysname{} with one layer QAOA and one iteration), which is already on the verge for practical quantum devices. In the near-term, quantum devices with about an estimated 3--15k qubits will meet the FEC computational demand of up to 50~MHz bandwidth 32-antenna base stations. While the required qubit count is within the scope of near-term devices, hybrid quantum--classical approaches may further advantage \sysname{}'s deployment in practical settings. Our studies inform NextG quantum devices and NextG wireless networks, and may motivate further research along exploiting temporal similarity from a native quantum hardware design perspective.

%% file: acks.tex
\section*{Acknowledgements}
This research is supported in part by the US Department of Energy Office of Advanced Scientific Computing Research, Accelerated Research for Quantum Computing Program. This research used resources of the National Energy Research Scientific Computing Center, a DOE Office of Science User Facility using NERSC award NERSC DDR-ERCAP0030278. This research is also supported by National Science Foundation (NSF) Award CNS-1824357.

%% file: asplos24.bbl
\begin{thebibliography}{10}

\bibitem{rappaport2013millimeter}
Theodore~S Rappaport, Shu Sun, Rimma Mayzus, Hang Zhao, Yaniv Azar, Kevin Wang, George~N Wong, Jocelyn~K Schulz, Mathew Samimi, and Felix Gutierrez.
\newblock Millimeter wave mobile communications for {5G} cellular: It will work!
\newblock {\em IEEE access}, 1:335--349, 2013.

\bibitem{bogale2016massive}
Tadilo~Endeshaw Bogale and Long~Bao Le.
\newblock Massive {MIMO} and mmwave for {5G} wireless hetnet: Potential benefits and challenges.
\newblock {\em IEEE Vehicular Technology Magazine}, 11(1):64--75, 2016.

\bibitem{kim2021heuristic}
Minsung Kim, Srikar Kasi, P~Aaron Lott, Davide Venturelli, John Kaewell, and Kyle Jamieson.
\newblock Heuristic quantum optimization for {6G} wireless communications.
\newblock {\em IEEE Network}, 35(4):8--15, 2021.

\bibitem{kasi2021cost}
Srikar Kasi, Paul Warburton, John Kaewell, and Kyle Jamieson.
\newblock A cost and power feasibility analysis of quantum annealing for {NextG} cellular wireless networks.
\newblock {\em IEEE Transactions on Quantum Engineering}, 4:1--17, 2023.

\bibitem{peterson1972error}
William~Wesley Peterson and Edward~J Weldon.
\newblock {\em Error-correcting codes}.
\newblock MIT press, 1972.

\bibitem{shannon1948mathematical}
Claude~Elwood Shannon.
\newblock A mathematical theory of communication.
\newblock {\em The Bell system technical journal}, 27(3):379--423, 1948.

\bibitem{itrs}
ITRS.
\newblock International technology roadmap for semiconductors 2.0, executive report, 2015.

\bibitem{xu2007complexity}
Weiyu Xu and Babak Hassibi.
\newblock On the complexity of exact maximum-likelihood decoding for asymptotically good low density parity check codes: A new perspective.
\newblock In {\em 2007 IEEE Information Theory Workshop}, pages 150--155. IEEE, 2007.

\bibitem{kasi2020towards}
Srikar Kasi and Kyle Jamieson.
\newblock Towards quantum belief propagation for {LDPC} decoding in wireless networks.
\newblock In {\em Proceedings of the 26th Annual International Conference on Mobile Computing and Networking}, pages 1--14, 2020.

\bibitem{kasi2022design}
Srikar Kasi, John Kaewell, and Kyle Jamieson.
\newblock The design and implementation of a hybrid classical-quantum annealing {Polar} decoder.
\newblock In {\em 2022 IEEE Global Communications Conference}, pages 5819--5825. IEEE, 2022.

\bibitem{bian2014discrete}
Zhengbing Bian, Fabian Chudak, Robert Israel, Brad Lackey, William~G Macready, and Aidan Roy.
\newblock Discrete optimization using quantum annealing on sparse ising models.
\newblock {\em Frontiers in Physics}, 2:56, 2014.

\bibitem{chancellor2016direct}
Nicholas Chancellor, Stefan Zohren, Paul~A Warburton, Simon~C Benjamin, and Stephen Roberts.
\newblock A direct mapping of max k-sat and high order parity checks to a chimera graph.
\newblock {\em Scientific reports}, 6(1):37107, 2016.

\bibitem{Shor_1997}
Peter~W. Shor.
\newblock Polynomial-time algorithms for prime factorization and discrete logarithms on a quantum computer.
\newblock {\em SIAM Journal on Computing}, 26(5):1484–1509, Oct 1997.

\bibitem{kandala2017hardware}
Abhinav Kandala, Antonio Mezzacapo, Kristan Temme, Maika Takita, Markus Brink, Jerry~M Chow, and Jay~M Gambetta.
\newblock Hardware-efficient variational quantum eigensolver for small molecules and quantum magnets.
\newblock {\em Nature}, 549(7671):242--246, 2017.

\bibitem{moll2018quantum}
Nikolaj Moll, Panagiotis Barkoutsos, Lev~S Bishop, Jerry~M Chow, Andrew Cross, Daniel~J Egger, Stefan Filipp, Andreas Fuhrer, Jay~M Gambetta, Marc Ganzhorn, et~al.
\newblock Quantum optimization using variational algorithms on near-term quantum devices.
\newblock {\em Quantum Science and Technology}, 3(3):030503, 2018.

\bibitem{biamonte2017quantum}
Jacob Biamonte, Peter Wittek, Nicola Pancotti, Patrick Rebentrost, Nathan Wiebe, and Seth Lloyd.
\newblock Quantum machine learning.
\newblock {\em Nature}, 549(7671):195--202, 2017.

\bibitem{preskill2018quantum}
John Preskill.
\newblock Quantum computing in the {NISQ} era and beyond.
\newblock {\em Quantum}, 2:79, 2018.

\bibitem{farhi2014quantum}
Edward Farhi, Jeffrey Goldstone, and Sam Gutmann.
\newblock A quantum approximate optimization algorithm, 2014.

\bibitem{gallager1962low}
Robert Gallager.
\newblock Low-density parity-check codes.
\newblock {\em IRE Transactions on information theory}, 8(1):21--28, 1962.

\bibitem{tanner1981recursive}
R~Tanner.
\newblock A recursive approach to low complexity codes.
\newblock {\em IEEE Transactions on information theory}, 27(5):533--547, 1981.

\bibitem{nguyen2019efficient}
Tram Thi~Bao Nguyen, Tuy Nguyen~Tan, and Hanho Lee.
\newblock Efficient {QC-LDPC} encoder for {5G} new radio.
\newblock {\em Electronics}, 8(6):668, 2019.

\bibitem{hailes2015survey}
Peter Hailes, Lei Xu, Robert~G Maunder, Bashir~M Al-Hashimi, and Lajos Hanzo.
\newblock A survey of {FPGA}-based {LDPC} decoders.
\newblock {\em IEEE Communications Surveys \& Tutorials}, 18(2):1098--1122, 2015.

\bibitem{10246149}
Srikar Kasi, John Kaewell, and Kyle Jamieson.
\newblock A quantum annealer-enabled decoder and hardware topology for {NextG} wireless {Polar} codes.
\newblock {\em IEEE Transactions on Wireless Communications}, 2023.

\bibitem{arikan2009channel}
Erdal Arikan.
\newblock Channel polarization: A method for constructing capacity-achieving codes for symmetric binary-input memoryless channels.
\newblock {\em IEEE Transactions on information Theory}, 55(7):3051--3073, 2009.

\bibitem{3gpp}
3rd Generation Partnership Project~(3GPP).
\newblock Multiplexing and channel coding.
\newblock {\em 38.212}, V.15.3.0, 2018.

\bibitem{bioglio2020design}
Valerio Bioglio, Carlo Condo, and Ingmar Land.
\newblock Design of {Polar} codes in {5G} new radio.
\newblock {\em IEEE Communications Surveys \& Tutorials}, 23(1):29--40, 2020.

\bibitem{tal2015list}
Ido Tal and Alexander Vardy.
\newblock List decoding of {Polar} codes.
\newblock {\em IEEE transactions on information theory}, 61(5):2213--2226, 2015.

\bibitem{ravi2022quantum}
Gokul~Subramanian Ravi, Kaitlin~N. Smith, Pranav Gokhale, and Frederic~T. Chong.
\newblock Quantum computing in the cloud: Analyzing job and machine characteristics, 2022.

\bibitem{O_Gorman_2017}
Joe O’Gorman and Earl~T. Campbell.
\newblock Quantum computation with realistic magic-state factories.
\newblock {\em Physical Review A}, 95(3), Mar 2017.

\bibitem{czarnik2020error}
Piotr Czarnik, Andrew Arrasmith, Patrick~J. Coles, and Lukasz Cincio.
\newblock Error mitigation with clifford quantum-circuit data, 2020.

\bibitem{Rosenberg2021}
Eliott Rosenberg, Paul Ginsparg, and Peter~L. McMahon.
\newblock Experimental error mitigation using linear rescaling for variational quantum eigensolving with up to 20 qubits.
\newblock {\em Quantum Science and Technology}, Nov 2021.

\bibitem{barron2020measurement}
George~S. Barron and Christopher~J. Wood.
\newblock Measurement error mitigation for variational quantum algorithms, 2020.

\bibitem{botelho2021error}
Ludmila Botelho, Adam Glos, Akash Kundu, Jarosław~Adam Miszczak, Özlem Salehi, and Zoltán Zimborás.
\newblock Error mitigation for variational quantum algorithms through mid-circuit measurements, 2021.

\bibitem{wang2021error}
Samson Wang, Piotr Czarnik, Andrew Arrasmith, M.~Cerezo, Lukasz Cincio, and Patrick~J. Coles.
\newblock Can error mitigation improve trainability of noisy variational quantum algorithms?, 2021.

\bibitem{takagi2021fundamental}
Ryuji Takagi, Suguru Endo, Shintaro Minagawa, and Mile Gu.
\newblock Fundamental limits of quantum error mitigation, 2021.

\bibitem{temme2017error}
Kristan Temme, Sergey Bravyi, and Jay~M Gambetta.
\newblock Error mitigation for short-depth quantum circuits.
\newblock {\em Physical review letters}, 119(18):180509, 2017.

\bibitem{li2017efficient}
Ying Li and Simon~C. Benjamin.
\newblock Efficient variational quantum simulator incorporating active error minimization.
\newblock {\em Phys. Rev. X}, 7:021050, Jun 2017.

\bibitem{giurgica2020digital}
Tudor Giurgica-Tiron, Yousef Hindy, Ryan LaRose, Andrea Mari, and William~J Zeng.
\newblock Digital zero noise extrapolation for quantum error mitigation.
\newblock In {\em 2020 IEEE International Conference on Quantum Computing and Engineering (QCE)}, pages 306--316. IEEE, 2020.

\bibitem{ding2020systematic}
Yongshan Ding, Pranav Gokhale, Sophia~Fuhui Lin, Richard Rines, Thomas Propson, and Frederic~T Chong.
\newblock Systematic crosstalk mitigation for superconducting qubits via frequency-aware compilation.
\newblock {\em arXiv preprint arXiv:2008.09503}, 2020.

\bibitem{smith2021error}
Kaitlin~N Smith, Gokul~Subramanian Ravi, Prakash Murali, Jonathan~M Baker, Nathan Earnest, Ali Javadi-Abhari, and Frederic~T Chong.
\newblock Error mitigation in quantum computers through instruction scheduling.
\newblock {\em arXiv preprint arXiv:2105.01760}, 2021.

\bibitem{kim2023evidence}
Youngseok Kim, Andrew Eddins, Sajant Anand, Ken~Xuan Wei, Ewout Van Den~Berg, Sami Rosenblatt, Hasan Nayfeh, Yantao Wu, Michael Zaletel, Kristan Temme, et~al.
\newblock Evidence for the utility of quantum computing before fault tolerance.
\newblock {\em Nature}, 618(7965):500--505, 2023.

\bibitem{cerezo21_vqa}
M.~Cerezo, Andrew Arrasmith, Ryan Babbush, Simon~C. Benjamin, Suguru Endo, Keisuke Fujii, Jarrod~R. McClean, Kosuke Mitarai, Xiao Yuan, Lukasz Cincio, and Patrick~J. Coles.
\newblock Variational {Quantum} {Algorithms}.
\newblock {\em Nature Reviews Physics}, 3(9):625--644, August 2021.

\bibitem{peruzzo2014variational}
Alberto Peruzzo, Jarrod McClean, Peter Shadbolt, Man-Hong Yung, Xiao-Qi Zhou, Peter~J Love, Al{\'a}n Aspuru-Guzik, and Jeremy~L O’brien.
\newblock A variational eigenvalue solver on a photonic quantum processor.
\newblock {\em Nature communications}, 5:4213, 2014.

\bibitem{glover19}
Fred Glover, Gary Kochenberger, and Yu~Du.
\newblock A {Tutorial} on {Formulating} and {Using} {QUBO} {Models}.
\newblock {\em arXiv}, November 2019.

\bibitem{wang18}
Zhihui Wang, Stuart Hadfield, Zhang Jiang, and Eleanor~G. Rieffel.
\newblock Quantum {Approximate} {Optimization} {Algorithm} for {MaxCut}: {A} {Fermionic} {View}.
\newblock {\em Physical Review A}, 97(2):022304, February 2018.

\bibitem{farhi22}
Edward Farhi, Jeffrey Goldstone, Sam Gutmann, and Leo Zhou.
\newblock The {Quantum} {Approximate} {Optimization} {Algorithm} and the {Sherrington}-{Kirkpatrick} {Model} at {Infinite} {Size}.
\newblock {\em Quantum}, 6:759, July 2022.

\bibitem{barak15}
Boaz Barak, Ankur Moitra, Ryan O'Donnell, Prasad Raghavendra, Oded Regev, David Steurer, Luca Trevisan, Aravindan Vijayaraghavan, David Witmer, and John Wright.
\newblock Beating the random assignment on constraint satisfaction problems of bounded degree.
\newblock {\em arXiv}, August 2015.

\bibitem{hastings19}
M.~B. Hastings.
\newblock Classical and {Quantum} {Bounded} {Depth} {Approximation} {Algorithms}.
\newblock {\em arXiv}, August 2019.

\bibitem{marwaha21}
Kunal Marwaha.
\newblock Local classical {MAX}-{CUT} algorithm outperforms \$p=2\$ {QAOA} on high-girth regular graphs.
\newblock {\em Quantum}, 5, April 2021.

\bibitem{chou22}
Chi-Ning Chou, Peter~J. Love, Juspreet~Singh Sandhu, and Jonathan Shi.
\newblock Limitations of {Local} {Quantum} {Algorithms} on {Random} {Max}-k-{XOR} and {Beyond}, February 2022.

\bibitem{lin16}
Cedric Yen-Yu Lin and Yechao Zhu.
\newblock Performance of {QAOA} on {Typical} {Instances} of {Constraint} {Satisfaction} {Problems} with {Bounded} {Degree}.
\newblock {\em arXiv}, January 2016.

\bibitem{farhi15}
Edward Farhi, Jeffrey Goldstone, and Sam Gutmann.
\newblock A {Quantum} {Approximate} {Optimization} {Algorithm} {Applied} to a {Bounded} {Occurrence} {Constraint} {Problem}.
\newblock {\em arXiv}, June 2015.

\bibitem{hadfield19}
Stuart Hadfield, Zhihui Wang, Bryan O’Gorman, Eleanor~G. Rieffel, Davide Venturelli, and Rupak Biswas.
\newblock From the {Quantum} {Approximate} {Optimization} {Algorithm} to a {Quantum} {Alternating} {Operator} {Ansatz}.
\newblock {\em Algorithms}, 12(2):34, February 2019.

\bibitem{streif20}
Michael Streif and Martin Leib.
\newblock Training the quantum approximate optimization algorithm without access to a quantum processing unit.
\newblock {\em IOP}, 5(3):034008, May 2020.
\newblock Publisher: IOP Publishing.

\bibitem{cafqa}
Gokul~Subramanian Ravi, Pranav Gokhale, Yi~Ding, William~M. Kirby, Kaitlin~N. Smith, Jonathan~M. Baker, Peter~J. Love, Henry Hoffmann, Kenneth~R. Brown, and Frederic~T. Chong.
\newblock Cafqa: Clifford ansatz for quantum accuracy, 2022.

\bibitem{brandao18}
Fernando G. S.~L. Brandao, Michael Broughton, Edward Farhi, Sam Gutmann, and Hartmut Neven.
\newblock For {Fixed} {Control} {Parameters} the {Quantum} {Approximate} {Optimization} {Algorithm}'s {Objective} {Function} {Value} {Concentrates} for {Typical} {Instances}.
\newblock {\em arXiv:1812.04170 [quant-ph]}, December 2018.

\bibitem{galda21}
Alexey Galda, Xiaoyuan Liu, Danylo Lykov, Yuri Alexeev, and Ilya Safro.
\newblock Transferability of optimal {QAOA} parameters between random graphs.
\newblock {\em arXiv:2106.07531 [quant-ph]}, June 2021.

\bibitem{shaydulin21}
Ruslan Shaydulin, Kunal Marwaha, Jonathan Wurtz, and Phillip~C. Lotshaw.
\newblock {QAOAKit}: {A} {Toolkit} for {Reproducible} {Study}, {Application}, and {Verification} of the {QAOA}.
\newblock {\em arXiv:2110.05555 [quant-ph]}, November 2021.

\bibitem{sud22}
James Sud, Stuart Hadfield, Eleanor Rieffel, Norm Tubman, and Tad Hogg.
\newblock A {Parameter} {Setting} {Heuristic} for the {Quantum} {Alternating} {Operator} {Ansatz}.
\newblock {\em arXiv}, November 2022.

\bibitem{9259985}
Wim Lavrijsen, Ana Tudor, Juliane Müller, Costin Iancu, and Wibe de~Jong.
\newblock Classical optimizers for noisy intermediate-scale quantum devices.
\newblock In {\em 2020 IEEE International Conference on Quantum Computing and Engineering (QCE)}, pages 267--277, 2020.

\bibitem{linke17}
Norbert~M. Linke, Dmitri Maslov, Martin Roetteler, Shantanu Debnath, Caroline Figgatt, Kevin~A. Landsman, Kenneth Wright, and Christopher Monroe.
\newblock Experimental comparison of two quantum computing architectures.
\newblock {\em Proceedings of the National Academy of Sciences}, 114(13):3305--3310, March 2017.

\bibitem{ibm21}
IBM.
\newblock The {IBM} {Quantum} heavy hex lattice, February 2021.

\bibitem{jin22}
Yuwei Jin, Jason Luo, Lucent Fong, Yanhao Chen, Ari~B. Hayes, Chi Zhang, Fei Hua, and Eddy~Z. Zhang.
\newblock A {Structured} {Method} for {Compilation} of {QAOA} {Circuits} in {Quantum} {Computing}.
\newblock {\em arXiv}, July 2022.

\bibitem{murali2019noise}
Prakash Murali, Jonathan~M Baker, Ali Javadi-Abhari, Frederic~T Chong, and Margaret Martonosi.
\newblock Noise-adaptive compiler mappings for noisy intermediate-scale quantum computers.
\newblock In {\em Proceedings of the Twenty-Fourth International Conference on Architectural Support for Programming Languages and Operating Systems}, pages 1015--1029, 2019.

\bibitem{pokharel2018demonstration}
Bibek Pokharel, Namit Anand, Benjamin Fortman, and Daniel~A Lidar.
\newblock Demonstration of fidelity improvement using dynamical decoupling with superconducting qubits.
\newblock {\em Physical review letters}, 121(22):220502, 2018.

\bibitem{jigsaw}
Poulami Das, Swamit Tannu, and Moinuddin Qureshi.
\newblock {JigSaw}: Boosting fidelity of {NISQ} programs via measurement subsetting.
\newblock In {\em MICRO-54: 54th Annual IEEE/ACM International Symposium on Microarchitecture}, MICRO '21, page 937–949, New York, NY, USA, 2021. Association for Computing Machinery.

\bibitem{murali2020software}
Prakash Murali, David~C McKay, Margaret Martonosi, and Ali Javadi-Abhari.
\newblock Software mitigation of crosstalk on noisy intermediate-scale quantum computers.
\newblock In {\em Proceedings of the Twenty-Fifth International Conference on Architectural Support for Programming Languages and Operating Systems}, pages 1001--1016, 2020.

\bibitem{ravi2021vaqem}
Gokul~Subramanian Ravi, Kaitlin~N. Smith, Pranav Gokhale, Andrea Mari, Nathan Earnest, Ali Javadi-Abhari, and Frederic~T. Chong.
\newblock Vaqem: A variational approach to quantum error mitigation, 2021.

\bibitem{ravi2022navigating}
Gokul~Subramanian Ravi, Kaitlin~N. Smith, Jonathan~M. Baker, Tejas Kannan, Nathan Earnest, Ali Javadi-Abhari, Henry Hoffmann, and Frederic~T. Chong.
\newblock Navigating the dynamic noise landscape of variational quantum algorithms with qismet, 2022.

\bibitem{dangwal2023varsaw}
Siddharth Dangwal, Gokul~Subramanian Ravi, Poulami Das, Kaitlin~N. Smith, Jonathan~M. Baker, and Frederic~T. Chong.
\newblock Varsaw: Application-tailored measurement error mitigation for variational quantum algorithms, 2023.

\bibitem{QMod}
Russfein.
\newblock Quantum computing modalities – a qubit primer revisited, October 2022.

\bibitem{Shi_2019}
Yunong Shi, Nelson Leung, Pranav Gokhale, Zane Rossi, David~I. Schuster, Henry Hoffmann, and Frederic~T. Chong.
\newblock Optimized compilation of aggregated instructions for realistic quantum computers.
\newblock {\em Proceedings of the Twenty-Fourth International Conference on Architectural Support for Programming Languages and Operating Systems}, Apr 2019.

\bibitem{Johansson_2012}
J.R. Johansson, P.D. Nation, and Franco Nori.
\newblock {QuTiP}: An open-source python framework for the dynamics of open quantum systems.
\newblock {\em Computer Physics Communications}, 183(8):1760--1772, aug 2012.

\bibitem{petersson2020discrete}
N.~Anders Petersson, Fortino~M. Garcia, Austin~E. Copeland, Ylva~L. Rydin, and Jonathan~L. DuBois.
\newblock Discrete adjoints for accurate numerical optimization with application to quantum control, 2020.

\bibitem{günther2021quandary}
Stefanie Günther, N.~Anders Petersson, and Jonathan~L. Dubois.
\newblock Quandary: An open-source c++ package for high-performance optimal control of open quantum systems, 2021.

\bibitem{qiskit23}
{Qiskit contributors}.
\newblock Qiskit: An open-source framework for quantum computing, 2023.

\bibitem{shaydulin23}
Ruslan Shaydulin, Phillip~C. Lotshaw, Jeffrey Larson, James Ostrowski, and Travis~S. Humble.
\newblock Parameter {Transfer} for {Quantum} {Approximate} {Optimization} of {Weighted} {MaxCut}.
\newblock {\em ACM Transactions on Quantum Computing}, 4, September 2023.

\bibitem{virtanen20}
Pauli Virtanen, Ralf Gommers, Travis~E. Oliphant, Matt Haberland, Tyler Reddy, David Cournapeau, Evgeni Burovski, Pearu Peterson, Warren Weckesser, Jonathan Bright, Stefan~J. van~der Walt, Matthew Brett, Joshua Wilson, K.~Jarrod Millman, Nikolay Mayorov, Andrew R.~J. Nelson, Eric Jones, Robert Kern, Eric Larson, C.~J. Carey, Ilhan Polat, Yu~Feng, Eric~W. Moore, Jake VanderPlas, Denis Laxalde, Josef Perktold, Robert Cimrman, Ian Henriksen, E.~A. Quintero, Charles~R. Harris, Anne~M. Archibald, Antonio~H. Ribeiro, Fabian Pedregosa, and Paul van Mulbregt.
\newblock {SciPy} 1.0: fundamental algorithms for scientific computing in {Python}.
\newblock {\em Nature Methods}, 17(3), March 2020.

\bibitem{mckay17}
David~C. McKay, Christopher~J. Wood, Sarah Sheldon, Jerry~M. Chow, and Jay~M. Gambetta.
\newblock Efficient {Z}-{Gates} for {Quantum} {Computing}.
\newblock {\em Physical Review A}, 96, August 2017.

\bibitem{glover1977heuristics}
Fred Glover.
\newblock Heuristics for integer programming using surrogate constraints.
\newblock {\em Decision sciences}, 8(1):156--166, 1977.

\bibitem{qbsolv}
{D-Wave Systems.} {Qbsolv Github Documentation}, 2021.

\bibitem{glover2010diversification}
Fred Glover, Zhipeng L{\"u}, and Jin-Kao Hao.
\newblock Diversification-driven tabu search for unconstrained binary quadratic problems.
\newblock {\em 4OR}, 8:239--253, 2010.

\bibitem{IBM-HW}
{IBM} quantum hardware roadmap.
\newblock \newline https://www.ibm.com/blogs/research/2020/09/ibm-quantum-roadmap/.

\bibitem{beverland2022assessing}
Michael~E. Beverland, Prakash Murali, Matthias Troyer, Krysta~M. Svore, Torsten Hoefler, Vadym Kliuchnikov, Guang~Hao Low, Mathias Soeken, Aarthi Sundaram, and Alexander Vaschillo.
\newblock Assessing requirements to scale to practical quantum advantage, 2022.

\end{thebibliography}
